\documentclass{nature}

\usepackage[utf8]{inputenc}
\usepackage{eso-pic}
\usepackage{graphicx}
\usepackage{color}
\usepackage{amsmath,amssymb,calc}
\usepackage{setspace}
\usepackage{ulem}
\usepackage{gensymb}
\usepackage{textcomp}
\usepackage{lineno}
\usepackage{hyperref}
\usepackage{newfloat}
\DeclareFloatingEnvironment[name={Supplementary Figure} ]{suppfigure}
\DeclareFloatingEnvironment[name={Supplementary Table} ]{supptable}

\usepackage{url}

\usepackage{soul}

\title{Optimizing carbon dioxide trapping for geological storage}

\author{Jaione Tirapu Azpiroz$^1$, Ronaldo Giro$^2$, Rodrigo Neumann Barros Ferreira$^1$, Marcio Nogueira Pereira da Silva$^1$,  Manuela Fernandes Blanco Rodriguez$^1$,  Adolfo E. Correa Lopez$^1$, David A. Lazo Vasquez$^3$, Matheus Esteves Ferreira$^1$, Mariana Del Grande$^1$, Ademir Ferreira Da Silva$^1$, Mathias B. Steiner$^1$}

\begin{document}

\maketitle

\begin{affiliations}
 \item IBM Research, Av. República do Chile, 330, CEP 20031-170, Rio de Janeiro, RJ, Brazil.
\item IBM Research, Rd J Fco Aguirre Proenca Km 9 Sp101, Hortolandia, Sao Paulo, 13186-900, SP, Brazil.
\item IBM Research, Rua Tutóia, 1157, São Paulo, Sao Paulo, 04007-005, SP, Brazil.
\end{affiliations}

\begin{abstract}

Carbon dioxide (CO$_2$) trapping in capillary networks of reservoir rocks is a pathway to long-term geological storage. At pore scale, the CO$_2$ trapping potential depends on injection pressure, temperature, and the rock's interaction with the surrounding fluids. Modeling this interaction requires adequate representations of both capillary volume and surface. For the lack of scalable representations, however, the prediction of a rock's CO$_2$ storage potential has been challenging. Here, we report how to represent a rock's pore space by statistically sampled capillary networks (ssCN) that preserve morphological rock characteristics. We have used the ssCN method to simulate CO$_2$ drainage within a representative sandstone sample at reservoir pressures and temperatures, exploring intermediate- and CO$_2$-wet conditions. This wetting regime is often neglected, despite evidence of plausibility. By raising pressure and temperature we observe increasing CO$_2$ penetration within the capillary network. For contact angles approaching 90$^\circ$, the CO$_2$ saturation exhibits a pronounced maximum reaching 80$\%$ of the accessible pore volume. This is about twice as high as the saturation values reported previously. For enabling validation of our results and a broader application of our methodology, we have made available the rock tomography data, the digital rock computational workflows, and the ssCN models used in this study.

\end{abstract}

\begin{introduction}

To achieve global targets for mitigating greenhouse gas emission, efficient carbon capture and storage technologies are needed.\cite{metz2005ipcc} Captured and purified CO$_2$ can be injected into subsurface rock formations, providing a pathway to long-term geological storage at scale. Encouragingly, estimates of the connected void space in reservoir rock of suitable geological formations exceed the volume required for satisfying the global CO$_2$ storage needs.\cite{kramer2020negative}
The physical and chemical processes underlying CO$_2$ geological storage  are: (i) structural storage under impermeable cap rocks; (ii) residual or capillary trapping; (iii) dissolution in water/brine and (iv) mineralization. Processes (i) and (ii) occur at shorter time scales - typically within a few years - but carry an elevated risk of CO$_2$ leakage\cite{Krevor2015}. Water-wet reservoir conditions lower the risk of CO$_2$ leakage in both structural and residual trapping, either by lowering the CO$_2$ permeability of the cap rock, or by raising the capillary pressures for increased CO$_2$ trapping efficiency at pore scale.\cite{CHIQUET2007,Saraji2013,Al-Khdheeawi2017} 
The processes (iii) and (iv) carry a lower risk of CO$_2$ leakage, however, they occur at longer time scales, ranging  from decades to centuries\cite{Krevor2015}. Capillary trapping of CO$_2$ has been reported to achieve saturation levels of 10-30$\%$ of the rock's pore volume\cite{Krevor2015} and is a promising candidate process for low-risk CO$_2$ storage at scale.

The efficiency of trapping CO$_2$ in deep geological formations is determined by the microscopic properties of the rock's connected pore space and depends on its interaction with supercritical CO$_2$ (scCO$_2$) and brine\cite{AlMenhali2016,Krevor2015,Krevor2012,Shi2011,Bennion2010,Bennion2008,Hu2017,Gooya2019,Kalam2020,Hinton2021,Guo2022}. Extensive research has been carried out for investigating the role of surface wettability on fluid displacement\cite{Guo2022}; the relative permeability and trapping of CO$_2$ in sandstone\cite{Krevor2012,Shi2011,Bennion2008}, as well as in carbonate\cite{Bennion2010,Bennion2008}, in shale and in anhydrite rocks\cite{Bennion2008}. In the case of sandstones, which are promising candidates for CO$_2$ storage at field scale\cite{Krevor2015}, experimental studies of CO$_2$ capillary trapping at reservoir conditions yielded maximum CO$_2$ saturation values of 46-59\% of rock pore volume\cite{Krevor2012}. 
CO$_2$ is injected into subsurface porous formations during drainage, displacing the resident fluid as it migrates withing the pore space as the non-wetting reservoir in response to the pressure gradients.\cite{Krevor2015, Alhosani2020} While strongly water-wet reservoirs are characteristic of residual CO$_2$ trapping, the maximum saturation achievable is limited.\cite{Krevor2012} Recent studies have, therefore, investigated the potential of intermediate- and CO$_2$-wet reservoir conditions\cite{PLUG2007, Saraji2013, CHIQUET2007}, including a wide range of wettability values from strongly water-wet to CO$_2$-wet.\cite{Iglauer2017}
However, for determining optimum CO$_2$ trapping conditions, the evaluation would benefit from controlled interrogation of the parameter space.  

Suitable numerical modeling approaches for simulating flow in reservoir rock at pore scale include mesh- or lattice-based direct simulation methods\cite{huang2005,Hu2017,Pan2004} for achieving high fidelity, and network based methods\cite{Valvatne2004,Raeini2017,Neumann2021} with improved computational efficiency. For optimizing the CO$_2$ capillary trapping in a given rock, it is necessary to screen for pressure and temperature conditions as a function of wettability, i.e., the contact angle at the interface between the rock, scCO$_2$ and brine. While prior studies have demonstrated the dependence of a rock's CO$_2$ storage potential on both the capillary network and its surface wetting properties\cite{Guo2022}, mapping the relevant parameter space with lab experiments in a controlled manner is impractical. 

In the following, we report a method designed to overcome these limitation by balancing accuracy and computational cost in simulations of liquid-solid interactions in capillary networks of reservoir rock. We apply the new method to simulate CO$_2$ trapping in sandstone under realistic reservoir conditions, tightly mapping the relevant parameter space for identifying optimum storage conditions.    

\end{introduction}

\begin{results}

\section*{Simplified Network Methodology}

We have developed a pore-scale flow simulator for studying injection and saturation of porous rock samples modeled as a network of capillaries\cite{Neumann2021,TirapuGHGT}. Specifically, we have simulated the injection of supercritical CO$_2$ into the capillary network model of a Berea Sister Gray sandstone sample filled with water as the resident fluid. Computationally, we track the displacement in time of the fluid interface between scCO$_2$ and water within each capillary of the connected pore space. 
To overcome the computational limitations, we have developed a sampling technique in which the assessment is performed with the aggregate result of multiple flow simulations performed with sets of much smaller but statistically equivalent capillary network models taken from the same digital rock sample. Importantly, each sub-sample matches statistically the morphlogical and geometrical properties of the original capillary network. 

In Fig.~\ref{Fig1_Methodology} we display the methodological workflow. As input, we have used grayscale rock images acquired with X-ray $\mu$CT scans\cite{lucas2022micro, Esteves2023}, and that we made publicly available as a digital rock dataset.\cite{NeumannDataset2020, Esteves_Ferreira2023}  A sequence of image processing steps are applied to the image for reducing noise, increasing contrast and for separating solid and void spaces (see \textit{Supplementary Information} section \ref{sec:SuppDigitalRock} and Supplementary Fig. \ref{DigitalRock}). As a representative rock sample, we have chosen a sandstone referred to as ``Berea Sister Gray''. We have used the Centerline algorithm to  create a CNM of the binarized digital rock image\cite{Neumann2021}. The CNM extracts from the pore space a voxel-wide line at the center of the pore channels, annotated with the pore radii at each point in space. The pore space is represented as a sequence of short cylinders with gradually changing radii. Each cylinder radius is defined as the radius of inscribed circle into the pore space centered at the point belonging to the line at the center of the pore channel (more details can be found in section \textit{Methods} and in section \ref{sec:CNM} of the \textit{Supplementary Information}). An example of the resulting CNM for a REV-sized sandstone sample is shown in Fig. \ref{Fig1_Methodology}. Supplementary Fig. \ref{PressureFields} displays the induced pressure field inside the network when an external 10 kPa/m pressure gradient is imposed along each axis.

\begin{figure}[!htb]
  \includegraphics[width=\linewidth]{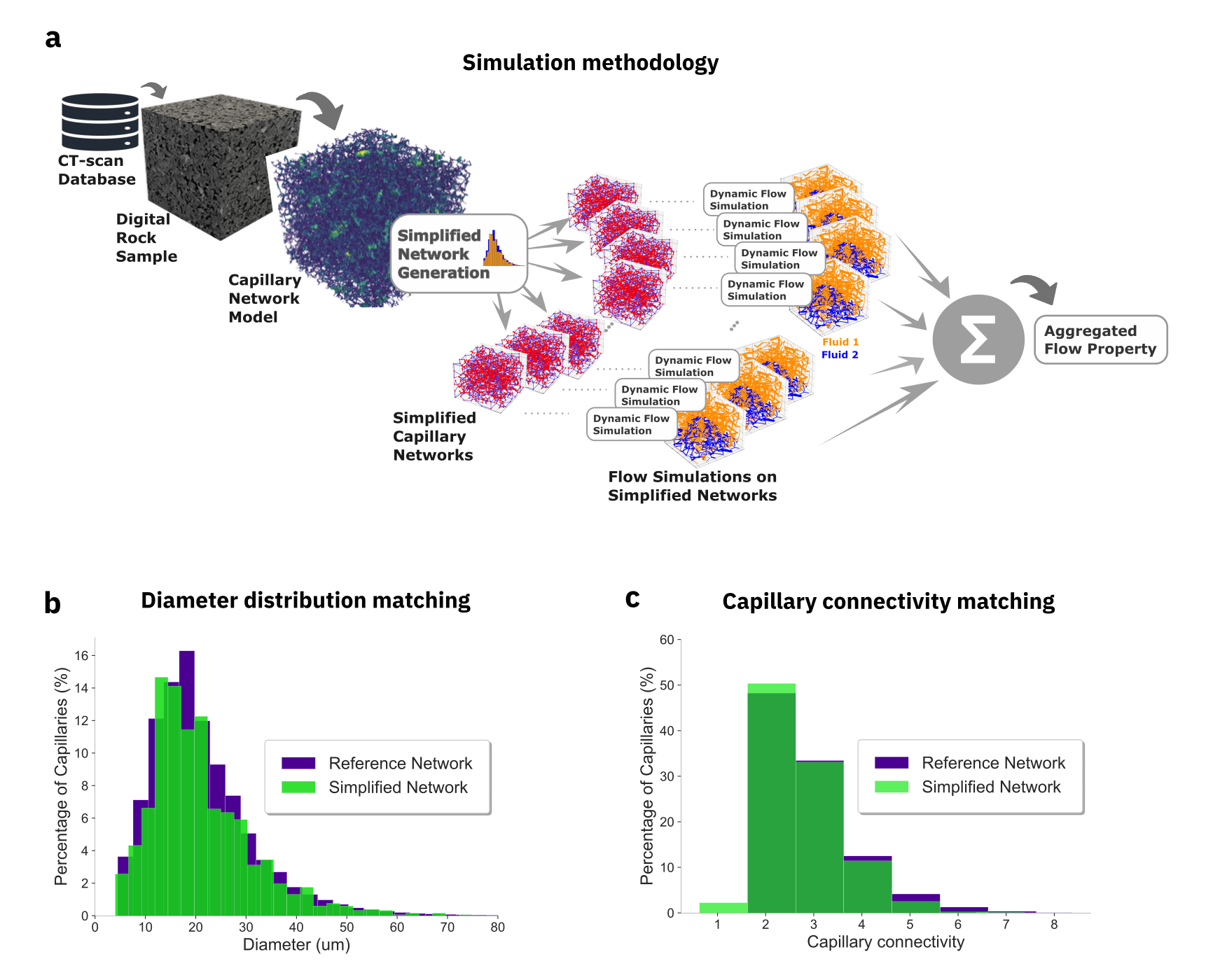}
  \caption{\textbf{Simplified network methodology.} a) Schematic workflow from rock tomography to flow properties. Distribution of b)  capillary diameter and c) capillary connectivity for the capillary network of the original rock sample as a reference and a statistically simplified one for comparison.}
  \label{Fig1_Methodology}
\end{figure}

We have built sets of statistically sampled capillary network (ssCN) representations preserving the following properties: (i) porosity; (ii) capillary diameter distribution; and (iii) node coordination number distribution. As a result, we obtain morphologically equivalent networks containing only a small fraction of the capillaries, i.e., $2.7 \times 10^6$ capillaries are reduced to $1.8 \times 10^3$ capillaries.
Each simplified network in the set begins as a random distribution of nodes and capillaries. Following an optimization routine that aims to match the properties of capillary connectivity, length, diameter, and general porosity of the original rock sample, it is then transformed into a network of capillaries with equivalent morphological properties as the original one. The result is an ensemble of simplified network models with significantly reduced number of capillaries that maintains the original CNM properties. 

This equivalence is seen by comparing the distributions of capillary diameters and  capillary connectivities displayed in Fig.~\ref{Fig1_Methodology}b and c, respectively, as percentage of the total number of capillaries between the original rock sample and the algorithmically generated one. The final step in our workflow consists of collecting the flow properties of the original rock sample from the average of the results taken from the set of $k$-ssCN simplified equivalent networks. We find that 1500 capillaries per ssCN, about $0.5\%$ of the original number, provides a good balance between computational accuracy and cost. As a result, the computed permeability averaged over an ensemble of 50 ssCN instances falls within $\pm 3 \sigma$ of the original value. Increasing the number of capillaries in the simplified representation reduces the variability of the results, however, it does not significantly improve the accuracy of the average estimate when compared to the original. 
More details can be found in Section \ref{SimplifiedCNMcubic} of the \textit{Supplementary Information}.

\section*{Simulation of CO$_2$ Capillary Trapping}

Following the method outlined above, we have performed a sensitivity analysis with respect to multiple fluid parameters, such as temperature, contact angle, pressure gradient, and quantified their influence on the infiltration and retention of CO$_2$ inside an ensemble of simplified capillary networks that is representative of the target Berea Sister Gray sandstone sample. To determine under which conditions CO$_2$ storage through capillary trapping is maximized, we have considered temperature scenarios from 323 K to 473 K, at 50K intervals. For each temperature scenario, we have studied a range of pressure gradients and contact angles (see Table \ref{ParameterTable}). Rock wettability shows a large variability from water-wet to CO$_2$-wet\cite{Iglauer2017}. Clean mineral surfaces, such as quartz, calcite, feldspar, and mica are water-wet (CO$_2$ contact angle ranges between 120$^{\circ}$ and 180$^{\circ}$) due to their hydrophilic character. In subsurface systems at reservoir conditions, the presence of organic matter is very likely. Typically, these surfaces were aged in crude oil or coal and are CO$_2$-wet or intermediate-wet, with CO$_2$ contact angles ranging from 10$^{\circ}$ to 110$^{\circ}$.\cite{Iglauer2017} In the following, we define the contact angle with respect to the injected fluid (scCO$_2$), which is the complementary of the angles normally defined in the literature\cite{Iglauer2017, Hu2017,CHIQUET2007}, see Table \ref{ContributionsParameterTable}. 

Another parameter with large variability is temperature. Low-temperature reservoirs range from 293 K to 363 K, intermediate-temperature reservoirs from 363 K to 423 K, and high-temperature reservoirs from 423 K to 573 K.\cite{ZARROUK201913} In our studies, we have considered temperatures ranging from 323 K to 473 K, see Table \ref{ParameterTable}, assumed a fixed absolute pressure of 10 MPa, and set a range of pressure gradients between 1$\times10^4$ and 1$\times10^7$ Pa/m for driving the flow. Parameters of viscosity for the resident and injected fluids as well as the interfacial tension between the fluids, necessary to perform two-phase flow simulations, were extracted from the literature\cite{Huber2009, HEIDARYAN2011, Bachu2009}. At the temperatures considered in this work, we have applied the correlating equation for the viscosity of water as extracted by Huber \textit{et al.}\cite{Huber2009} in terms of the product of a temperature-dependent zero-density limit term, and a residual viscosity term that depends of both temperature and density as its value increases. The density of water at each value of temperature and pressure was calculated according to Wagner \textit{et al.}\cite{Wagner2002}. We have computed the viscosity of CO$_2$ under supercritical conditions of pressure and temperature following the correlation extracted by Heidaryan \textit{et al.}\cite{HEIDARYAN2011}. Finally, we have deduced the interfacial tension between supercritical CO$_2$ and water at the reservoir pressure for the various temperature scenarios from the work of Bachu and Bennion\cite{Bachu2009} as provided in Table \ref{ParameterTable}. 

\begin{figure}[!htb]
  \includegraphics[width=\linewidth]{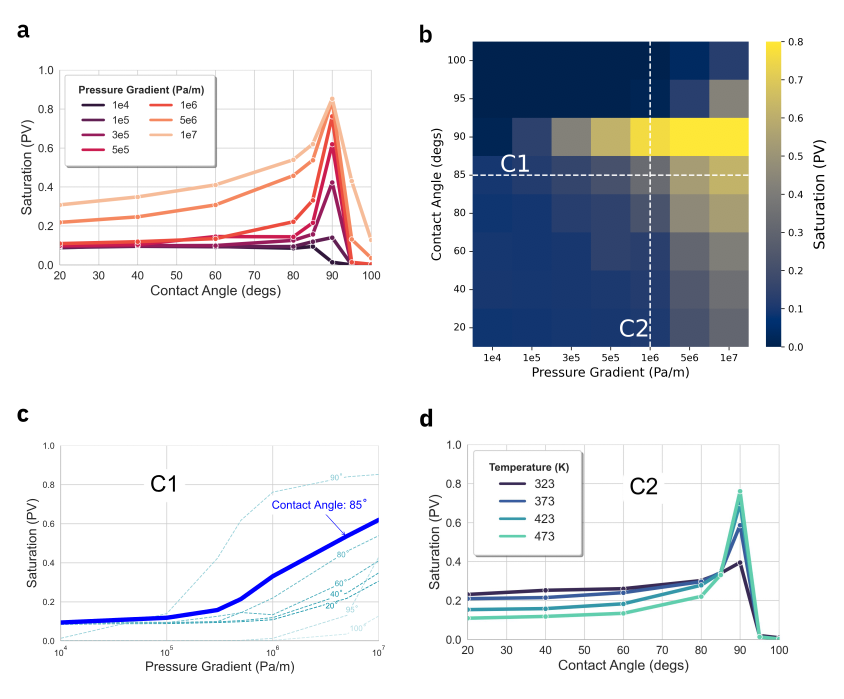}
  \caption{ \textbf{CO$_2$ saturation as function of pressure and contact angle.} a) Maximum {CO}$_2$ saturation as function of contact angle for representative pressure gradients at a temperature of 473 K. b) Distributional map of the maximum CO$_2$ saturation as a function of applied pressure gradient and contact angle, respectively, at a temperature of 473 K. c) {CO}$_2$ saturation along $C1$ cutline at a contact angle of 85$^\circ$. Dashed lines represent parallel cutlines to $C1$ taken at representative contact angles. d) {CO}$_2$ saturation along $C2$ cutline at a pressure gradient of 1$\times 10^6$ Pa/m for representative temperatures.} 
  \label{Fig2_SatCutlines}
\end{figure}

Fig. \ref{Fig2_SatCutlines}a shows the maximum scCO$_2$ saturation curves at T = 473 K plotted as function of fluid interface contact angle. The maximum scCO$_2$ saturation was extracted from the saturation as a function of time, as shown in Fig. \ref{SuppSatvsInjVol}. Each point in Fig. \ref{Fig2_SatCutlines}a represents the maximum value of CO$_2$ saturation from the average across 150 simulations (set of 50 simplified capillary networks along X, Y and Z axes) per set of injection conditions. 
Contact angles between 20$^{\circ}$ and 80$^{\circ}$ represent CO$_2$-wet regime where higher saturation by the injected fluid would be expected as capillary pressure favors the displacement of the resident fluid, particularly when, as per the calculations in Table \ref{ContributionsTable}, viscous forces remain lower than the externally applied. We observe that for pressure gradients below 1$\times10^5$ Pa/m, the maximum CO$_2$ saturation is around 15$\%$. This saturation value is almost constant for contact angles ranging from 20$^{\circ}$ to 80$^{\circ}$ (CO$_2$ wet range), with a slight increase around 85$^{\circ}$ for pressure gradient 1$\times10^5$ Pa/m. For larger contact angles (water-wet), the CO$_2$ saturation diminishes to zero. 
We note that the CO$_2$ saturation behaviour is a result of the collective interactions between all the pores in the capillary network and reflects the competing interplay between the externally applied pressure gradient and the balance of capillary pressures and viscous forces within each capillary of the network, see Eq. \ref{equation:twophase}. While the observed saturation behaviour is surprising and might appear counter-intuitive, it emerges from the complex interplay of capillaries in the network and indicates that pore scale analysis may reveal unexpected phenomena at larger scales. For externally applied pressure gradients above 1$\times10^5$ Pa/m, we observe in Fig. \ref{Fig2_SatCutlines}a a strong dependence of CO$_2$ saturation on contact angle. In particular, saturation of CO$_2$ increases monotonically as function of contact angle within the intermediate-wetting regime\cite{Iglauer2015} until reaching a peak around 90$^{\circ}$. Beyond that angle, CO$_2$ saturation rapidly decreases with the contact angle as the surface wettability turns to water wet.

In the regime of pressure gradients below 1$\times10^5$ Pa/m, fluid flow is mostly driven by capillary pressure. Applying the second term of the right hand side of Eq. \ref{equation:twophase} to the range of capillary diameters shown in Fig. \ref{Fig1_Methodology}b, we estimate the capillary pressure to be of the order of 4 kPa for contact angles between 20$^\circ$ and 80$^\circ$, the same order of magnitude as the simulated pressure distributions on capillary nodes shown in Fig. \ref{SuppPressuresCombined}a. 
When the fluid-solid interface contact angle $\theta$ approaches 90$^\circ$ degrees, capillary pressure drops to zero, see Eq. \ref{equation:twophase}. In this pressure regime, CO$_2$ saturation diminishes as the driving force is not strong enough for sustaining fluid flow. At low externally applied pressure gradients, the magnitude of the pressure within the capillary network is nearly negligible, see Fig. \ref{SuppPressuresCombined}b. The relatively low CO$_2$ saturation levels observed for some contact angles are likely due to the network complexity and the remaining viscous forces, with capillaries being plugged by interfaces of water/CO$_2$, where the capillary pressure counteracts fluid flow. 

The higher CO$_2$ saturation seen in the regime of pressure gradients above 1$\times10^5$ Pa/m stems from the fact that the pressure gradients induced by the external driving force are strong enough to overcome the counteracting capillary pressure on those capillaries that had become plugged under conditions of low contact angles, and thus enabling reaching significantly higher saturation values. Saturation values of 20$\%$ to 40$\%$ are observed for gradient pressure of 5$\times 10^6$ and 1$\times10^7$ Pa/m, respectively - see Fig. \ref{Fig2_SatCutlines}a. In Fig. \ref{SuppPressuresCombined}a, we observe  nodes with pressure of the same order of magnitude as the capillary pressure. Moreover, within the intermediate-wet regime, as the contact angle approaches 90$^\circ$, CO$_2$ saturation reaches a peak because capillary pressure drops to zero and fluid flow is driven primarily by the applied pressure gradient. Within this regime, the saturation values of CO$_2$ decrease rapidly with contact angles larger than 90$^\circ$, because the imposed pressure gradient is no longer high enough to counteract capillary pressure.

Fig. \ref{Fig2_SatCutlines}b maps the maximum achievable CO$_2$ saturation distribution with regards to applied pressure gradient and of CO$_2$-water-rock contact angle. The distinct region of sharp saturation increase around the 90$^\circ$ can be seen in Fig. \ref{Fig2_SatCutlines}c-d along the 2 cross-sectional lines indicated in Fig. \ref{Fig2_SatCutlines}b as $C1$ and $C2$, respectively.  As a key result of our study, we obtain a maximum of 85.6\% CO$_2$ saturation for a pressure gradient of $10\times10^7$ Pa/m and $T=473\,\text{K}$ using $90^\circ$ contact angle, for the representative sandstone sample.
In the cross-section plotted in Fig. \ref{Fig2_SatCutlines}c, we observe that the saturation increases with the applied pressure until it reaches a plateau. Variability of the saturation as a function of temperature is generally small as can be seen in Fig. \ref{Fig2_SatCutlines}d. The increased saturation following deeper permeation of the rock space by scCO$_2$ during injection within the intermediate-wet regime is consistent with a lower capillary pressure.\cite{Hu2017}. In the following, we will analyze the injection conditions for CO$_2$ storage in view of storage security and process efficiency.

\section*{Optimization of CO$_2$ Storage Conditions}

CO$_2$ storage security and costs are primary concerns when optimizing injection conditions. Injected fluid that is not trapped within the pore space will escape, as indicated by injected volumes larger than 1 pore volume (PV). Fig. \ref{Fig3_InjectionConditions}a sheds light on the relative amounts of trapped \textit{vs.} mobile scCO$_2$ within the sample by plotting the value of saturation as a function of the injected volume. We observe in Fig. \ref{Fig3_InjectionConditions}a that intermediate-wet conditions lead to the largest saturation values, especially around a contact angle of 90$^\circ$ and at higher pressures. However, this often requires injection of volumes larger than 1 PV. While the exact values depend on the injection conditions, the results suggest that intermediate-wet conditions may require lower injection pressures to enhance storage security. At lower temperatures, we observe a reduced level of saturation per injected volume, see Fig. \ref{MaxSatvsinjVolvsTemp}, requiring larger injected volumes. 

Operational efficiency of the process requires maximizing the volume of CO$_2$ stored while minimizing the volume of CO$_2$ injected, thus minimizing cost. The fraction of the injected CO$_2$ that passes through the pore space without being trapped leads to the saturation curve of scCO$_2$ forming a plateau as a function of time, as in Fig. \ref{SuppSatvsInjVol}a. This effect can also be analyzed as function of injected volume, see Fig. \ref{SuppSatvsInjVol}b, for the same simulation conditions. For the purpose of our analysis, we define the variable \textit{Weighted Saturation} ($wS$) as the CO$_2$ saturation (\textit{S}) scaled by the ratio of saturation to injected volume (\textit{IV}), that is, $wS = S\frac{S}{IV}$, with units of pore volumes of injected scCO$_2$ between $0$ and $1$. The peak value of $wS$ is reached for a value of injected volume below 1, as any additional injection does not further increase the saturation level. In Fig. \ref{SuppSatvsInjVol}c, we plot the weighted saturation as a function of injected volume for a range of pressures at a contact angle of $85^\circ$, and Fig. \ref{SuppSatvsInjVol}d displays the weighted saturation for a range of contact angles at a fixed 5$\times 10^6$ Pa/m applied pressure gradient. The plot in Fig. \ref{Fig3_InjectionConditions}b represents maximum weighted saturation relative to injected volume across all the simulated scenarios aimed at optimizing saturation close to maximum injection utilization. For contact angles around 90$^\circ$ and high applied pressures, we obtain the highest saturation values. However, lower applied pressures in the intermediate-wet regime like $90^\circ$ and $3 \times 10^5 - 5 \times 10^5$ Pa/m, might offer a better balance of safety and efficiency. Overall, higher temperatures seem to improve efficiency as well as safety,  see Fig. \ref{MaxSatvsinjVolvsTemp} and Fig. \ref{weightedSatvsTemp}. 

\begin{figure}[!htb]
  \includegraphics[width=\linewidth]{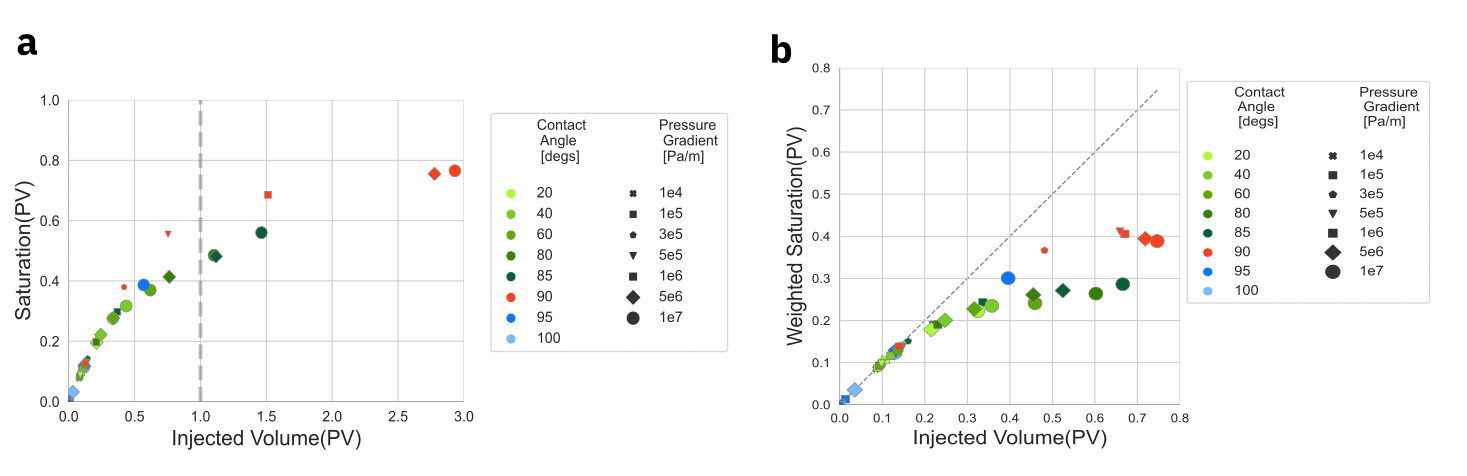}
  \caption{\textbf{Efficiency and security of CO$_2$ storage in capillary trapping}. Green color shades represent the CO$_2$-wet regime,  blue shades correspond to the water-wet regime, and red shades identify contact angles close to 90$^\circ$. The temperature is set to 473 K. Larger symbols represent higher pressure gradients. 
  a) CO$_2$ saturation at 90$\%$ of the maximum value as a function of the injected volume required to reach that value. Injected volume larger than 1 PV represents scCO$_2$ that is not stored within the sample.
  b) Relation of maximum weighted CO$_2$ saturation ($wS$) to injected volume across simulated conditions, aimed at increasing saturation close to the diagonal for maximum injection utilization.}
  \label{Fig3_InjectionConditions}
\end{figure}

In summary, we have developed a methodology to perform two-phase flow simulations in porous rock samples for exploring the drainage of supercritical carbon dioxide in brine-flooded sandstone reservoirs under intermediate- and CO$_2$-wet conditions.
We have analyzed a wide range of injection scenarios by varying contact angles, temperatures, and pressure gradients for identifying the best conditions for CO$_2$ storage through capillary trapping. Our CO$_2$ saturation results for contact angles between 20$^\circ$ and 40$^\circ$ are comparable with Krevor \textit{et al.}\cite{Krevor2015} for water-wet surfaces, where trapped saturation can reach up to 30\% of the rock's pore volume. For contact angles approaching the intermediate-wet regime around around 90$^\circ$, however, the CO$_2$ saturation efficiency increases sharply. For optimized conditions, we obtain saturation values as high as 80\% of the pore volume. Our results suggest that a high saturation efficiency may be maintained even at lower pressure levels which would significantly reduce the risk of CO$_2$ leakage.

\end{results}

\begin{methods}\label{sec:methods}

\subsection{Digital Rock Image processing} 

This work relies on high-resolution digital rock images of rock samples as the  geometrical basis for the pore scale simulations and the study of rock properties and fluid infiltration and storage within sub-surface porous structures\cite{ANDRA201333,Neumann2021}. In our work, we have acquired a large dataset of three-dimensional images extracted from X-ray computed micro-tomography ($\mu$CT) \cite{Neumann2021,Esteves2023} of sandstone and carbonate rock samples. The reconstructed volume of the fully digitized rock sample obtained from the $\mu$CT measurements is usually cropped into cube-shaped volumes that are more computationally manageable while retaining statistically significant rock properties. In our simulations, we employed a $(2.25\,\text{mm})^3$ digitized sandstone rock sample scanned at a resolution of $2.25\,\mu\text{m/voxel}$. This resulted in a digital sample with $1000^3$ voxels that was determined as the Representative Elementary Volume (REV) and therefore sufficient to yield accurate rock property predictions like porosity and permeability\cite{Neumann2021}. More details can be found in the Supplementary Information section \ref{sec:SuppDigitalRock}.

\subsection{Capillary Network extraction and centerlines representation}

We use a graph-based, capillary network model (CNM) as an accurate representation of the rock porous geometry. Starting with the 3D binarized digital rock image, we apply a custom Dijkstra’s Minimum Path algorithm\cite{Dijkstra59}, transforming the pore space into voxel-wide lines at the center of the pore channels, finding the most central path from inlet pore to outlet pore of each set of connected pore space voxels through a centrality-based cost function. Each section of the line is saved as a node in a graph and converted into a short cylinder with spatially varying radius to match the local geometry extracted from the microtomography. The resulting network of connected capillaries is an accurate representation of the rock's pore space. As an example of this transformation, we show the small $100^3$ image with $(225\,\mu\text{m})^3$ volume and its corresponding CNM in Fig. \ref{Centerlines} containing 4069 capillaries with a color-coded diameter scale. The capillary network of the sandstone sample analyzed in this paper displays a porosity of $33\%$ and results in a CNM with $2.7 \times 10^6$ capillaries. More details can be found in the Supplementary Information section \ref{sec:CNM}, and the full algorithm description can be found in the references\cite{Neumann2021}.

\subsection{Single-phase flow simulation}

The fined-grained capillary network representation of the rock's pore spatial distribution described in the previous section was employed to simulate both single and two-phase fluid flow with a high level of geometrical accuracy. 
We assume laminar flow and apply the equations relating pressure and flow within each capillary, followed by conservation of mass at each network node, to build a large system of coupled equations in sparse matrix form. 
The Hagen-Poiseuille equation is applied when simulating single-phase stationary flow, together with the conservation of mass at each network node, to build a large but rather sparse system of equations to extract properties like pressure distribution or flow rate at each point in the network of microscopic capillaries, as well as bulk flow properties like permeability \cite{Neumann2021}.
Fig. \ref{PressureFields} displays the results of running single-phase, pressure-driven, Poiseuille flow simulations on the capillary network representation of a $1000^3$ voxel sandstone rock sample as input geometry. The absolute permeability for this REV was computed from these simulations as 105, 92 and 54 mD along the X-, Y- and Z-axis, respectively. Further description can be found in the Supplementary Information section \ref{sec:SinglePhaseFlow}.

\subsection{Two-phase flow simulations}

Multi-phase flow simulations track the displacement in time of the fluid-fluid interface within each capillary of the high-resolution 3D geometric representation of the rock sample. We restrict the modeling to two incompressible fluids, under laminar, piston-like, one-dimensional flow along the length of each capillary. The pressure gradient between the two ends of each capillary is expressed as the sum of the gradients produced by various physical effects acting on the position $x(t)$ of the effective interface, after removing all annulling interfaces, between fluids. 
In our current implementation, we retain the effect of viscous forces and first order capillary pressure, after higher order dynamical effects to the interface shape are dismissed. 

Tracking of the fluid interfaces across the network of capillaries then proceeds in alternating sequences of \textit{free evolution} and \textit{jumps}. Free evolution refer to the time interval in which the interfaces progress along the same capillary and the overall number of interfaces within all capillaries in the network remain constant. 
Jump steps occur when any interface reaches a node, which means that the interface is ready to leave its current capillary and enter one or more new capillaries. In this step, free evolution pauses, the interfaces are redistributed throughout the network, and the system of differential algebraic equations representing the network dynamics is rewritten to account for the changes in interface locations. 
The simulator computes the velocity of the fluid flow through all capillaries as a function of time, which is integrated to determine the position of the fluid-fluid interface within each capillary at each time step, from which to deduce the level of saturation of each fluid phase in the rock pore space. More details on the formulation and implementation can be found in section \ref{sec:TwoPhaseFlow} of the Supplementary Information.

\subsection{Simplified capillary network representation}

Unlike single phase flow that can run within minutes even on the high-resolution capillary network representation of a REV-sized rock sample with millions of nodes and links, the two-phase dynamic simulations become unfeasible even on large computing resources. 
To overcome this limitation, we employ a smaller capillary network yet representative of the original rock full centerlines representation to run two-phase simulations and extract flow properties.
A custom algorithm is used to construct simplified 2D and 3D capillary networks capable of preserving geometrical properties of the rock morphology relevant to fluid flow such as : (i) porosity; (ii) capillary diameter distribution; and (iii) node coordination number distribution\cite{Raoof2009}. More details on the algorithm to synthesize networks can be found in section \ref{sec:SimplifiedCNM} of the Supplementary Information.

\subsection{Simulation Toolkit for Scientific Discovery (ST4SD):} In our study we scanned through 4 different temperature scenarios, and per scenario, we studied about 8 fluid-rock interface contact angle values and no less than 8 different driving pressure gradient cases per angle, totaling in 256 different injection conditions to be simulated. Per case simulated, flow simulations of a full data-set of 50 simplified capillary networks, assuming driving pressure along all three axes, that is, 150 executions per each of the 256 cases considered, requiring proper parsing and aggregation of nearly $4 \times 10^4$ simulations.  
In this work, we employed the Simulation Toolkit for Scientific Discovery (ST4SD)~\cite{ST4SD} to automate the execution of long simulation campaigns with several chained steps. The use of such workflow scheduler ensures the reproducibility of our results and enable efficiency gains by optimising the use of computing resources.
Fig. \ref{Fig1_Methodology} illustrates the conceptual workflow and in Fig. \ref{ST4SD}, we show the sequence of steps executed as an ST4SD experiment. A CNM representation of a rock sample is used as input to the ST4SD routine. 
In this workflow we generate tens to hundreds of simplified capillary networks that meaningfully represent the properties of the original rock sample network model (see section \textit{Simplified capillary network representation}). Each simplified capillary network is then used as geometrical input to parallel flow simulations that will estimate relevant physical properties in each representative system. Finally, the individual results from each simplified network are aggregated and combined into a single estimate that applies to the original network. See Supplementary Information section \ref{sec:ST4SD} for more details on this workflow.

\end{methods}

\begin{data}

Microtomography datasets containing grayscale and binary rock image data are available at the Digital Rocks Portal (\url{https://dx.doi.org/10.17612/f4h1-w124}) for sandstone samples and on Figshare (\url{https://doi.org/10.25452/figshare.plus.21375565}) for sandstone and carbonate samples. 

\end{data}

\begin{code}
The source code used to extract the CNM representation from the rock $\mu$CT image is available at \url{https://github.com/IBM/flowdiscovery-digital-rock}. Additional algorithms used for processing and segmenting the raw grayscale images, are available as Python code at \url{https://github.com/IBM/microCT-Dataset}. The code used to run single-phase and two-phase flow simulations on CNM representations of the rock samples is available at \url{https://github.com/IBM/flowdiscovery-simulator}. The code employed in the automation of the scientific workflows allowing the execution of ST4SD~\cite{ST4SD} experiments is available at \url{https://github.com/st4sd/flow-simulator-experiment}.

\end{code}


\begin{thebibliography}{10}
\expandafter\ifx\csname url\endcsname\relax
  \def\url#1{\texttt{#1}}\fi
\expandafter\ifx\csname urlprefix\endcsname\relax\def\urlprefix{URL }\fi
\providecommand{\bibinfo}[2]{#2}
\providecommand{\eprint}[2][]{\url{#2}}

\bibitem{metz2005ipcc}
\bibinfo{author}{Metz, B.}, \bibinfo{author}{Davidson, O.},
  \bibinfo{author}{De~Coninck, H.}, \bibinfo{author}{Loos, M.} \&
  \bibinfo{author}{Meyer, L.}
\newblock \emph{\bibinfo{title}{IPCC special report on carbon dioxide capture
  and storage}} (\bibinfo{publisher}{Cambridge: Cambridge University Press},
  \bibinfo{year}{2005}).

\bibitem{kramer2020negative}
\bibinfo{author}{Kramer, D.}
\newblock \bibinfo{title}{Negative carbon dioxide emissions}.
\newblock \emph{\bibinfo{journal}{Physics today}}
  \textbf{\bibinfo{volume}{73}}, \bibinfo{pages}{44--51}
  (\bibinfo{year}{2020}).

\bibitem{Krevor2015}
\bibinfo{author}{Krevor, S.} \emph{et~al.}
\newblock \bibinfo{title}{Capillary trapping for geologic carbon dioxide
  storage {\textendash} from pore scale physics to field scale implications}.
\newblock \emph{\bibinfo{journal}{International Journal of Greenhouse Gas
  Control}} \textbf{\bibinfo{volume}{40}}, \bibinfo{pages}{221--237}
  (\bibinfo{year}{2015}).
\newblock \urlprefix\url{https://doi.org/10.1016/j.ijggc.2015.04.006}.

\bibitem{CHIQUET2007}
\bibinfo{author}{CHIQUET, P.}, \bibinfo{author}{BROSETA, D.} \&
  \bibinfo{author}{THIBEAU, S.}
\newblock \bibinfo{title}{Wettability alteration of caprock minerals by carbon
  dioxide}.
\newblock \emph{\bibinfo{journal}{Geofluids}} \textbf{\bibinfo{volume}{7}},
  \bibinfo{pages}{112--122} (\bibinfo{year}{2007}).
\newblock
  \urlprefix\url{https://onlinelibrary.wiley.com/doi/abs/10.1111/j.1468-8123.2007.00168.x}.

\bibitem{Saraji2013}
\bibinfo{author}{Saraji, S.}, \bibinfo{author}{Goual, L.},
  \bibinfo{author}{Piri, M.} \& \bibinfo{author}{Plancher, H.}
\newblock \bibinfo{title}{Wettability of supercritical carbon
  dioxide/water/quartz systems: Simultaneous measurement of contact angle and
  interfacial tension at reservoir conditions}.
\newblock \emph{\bibinfo{journal}{Langmuir}} \textbf{\bibinfo{volume}{29}},
  \bibinfo{pages}{6856--6866} (\bibinfo{year}{2013}).
\newblock \urlprefix\url{https://doi.org/10.1021/la3050863}.
\newblock \bibinfo{note}{PMID: 23627310}.

\bibitem{Al-Khdheeawi2017}
\bibinfo{author}{Al-Khdheeawi, E.~A.}, \bibinfo{author}{Vialle, S.},
  \bibinfo{author}{Barifcani, A.}, \bibinfo{author}{Sarmadivaleh, M.} \&
  \bibinfo{author}{Iglauer, S.}
\newblock \bibinfo{title}{Influence of co2-wettability on co2 migration and
  trapping capacity in deep saline aquifers}.
\newblock \emph{\bibinfo{journal}{Greenhouse Gases: Science and Technology}}
  \textbf{\bibinfo{volume}{7}}, \bibinfo{pages}{328--338}
  (\bibinfo{year}{2017}).
\newblock
  \urlprefix\url{https://onlinelibrary.wiley.com/doi/abs/10.1002/ghg.1648}.

\bibitem{AlMenhali2016}
\bibinfo{author}{Al-Menhali, A.~S.} \& \bibinfo{author}{Krevor, S.}
\newblock \bibinfo{title}{Capillary trapping of {CO}$_2$ in oil reservoirs:
  Observations in a mixed-wet carbonate rock}.
\newblock \emph{\bibinfo{journal}{Environmental Science {\&} Technology}}
  \textbf{\bibinfo{volume}{50}}, \bibinfo{pages}{2727--2734}
  (\bibinfo{year}{2016}).
\newblock \urlprefix\url{https://doi.org/10.1021/acs.est.5b05925}.

\bibitem{Krevor2012}
\bibinfo{author}{Krevor, S. C.~M.}, \bibinfo{author}{Pini, R.},
  \bibinfo{author}{Zuo, L.} \& \bibinfo{author}{Benson, S.~M.}
\newblock \bibinfo{title}{Relative permeability and trapping of {CO}$_2$ and
  water in sandstone rocks at reservoir conditions}.
\newblock \emph{\bibinfo{journal}{Water Resources Research}}
  \textbf{\bibinfo{volume}{48}} (\bibinfo{year}{2012}).
\newblock \urlprefix\url{https://doi.org/10.1029/2011wr010859}.

\bibitem{Shi2011}
\bibinfo{author}{Shi, J.-Q.}, \bibinfo{author}{Xue, Z.} \&
  \bibinfo{author}{Durucan, S.}
\newblock \bibinfo{title}{Supercritical {CO}$_2$ core flooding and imbibition
  in berea sandstone {\textemdash} {CT} imaging and numerical simulation}.
\newblock \emph{\bibinfo{journal}{Energy Procedia}}
  \textbf{\bibinfo{volume}{4}}, \bibinfo{pages}{5001--5008}
  (\bibinfo{year}{2011}).
\newblock \urlprefix\url{https://doi.org/10.1016/j.egypro.2011.02.471}.

\bibitem{Bennion2010}
\bibinfo{author}{Bennion, D.~B.} \& \bibinfo{author}{Bachu, S.}
\newblock \bibinfo{title}{Drainage and imbibition {CO}$_2$/brine relative
  permeability curves at reservoir conditions for carbonate formations}.
\newblock In \emph{\bibinfo{booktitle}{All Days}} (\bibinfo{publisher}{{SPE}},
  \bibinfo{year}{2010}).
\newblock \urlprefix\url{https://doi.org/10.2118/134028-ms}.

\bibitem{Bennion2008}
\bibinfo{author}{Bennion, D.~B.} \& \bibinfo{author}{Bachu, S.}
\newblock \bibinfo{title}{Drainage and imbibition relative permeability
  relationships for supercritical {CO}$_2$/brine and {H}$_2${S}/brine systems
  in intergranular sandstone, carbonate, shale, and anhydrite rocks}.
\newblock \emph{\bibinfo{journal}{{SPE} Reservoir Evaluation {\&} Engineering}}
  \textbf{\bibinfo{volume}{11}}, \bibinfo{pages}{487--496}
  (\bibinfo{year}{2008}).
\newblock \urlprefix\url{https://doi.org/10.2118/99326-pa}.

\bibitem{Hu2017}
\bibinfo{author}{Hu, R.}, \bibinfo{author}{Wan, J.}, \bibinfo{author}{Kim, Y.}
  \& \bibinfo{author}{Tokunaga, T.~K.}
\newblock \bibinfo{title}{Wettability effects on supercritical
  {CO}$_2${\textendash}brine immiscible displacement during drainage:
  Pore-scale observation and {3D} simulation}.
\newblock \emph{\bibinfo{journal}{International Journal of Greenhouse Gas
  Control}} \textbf{\bibinfo{volume}{60}}, \bibinfo{pages}{129--139}
  (\bibinfo{year}{2017}).
\newblock \urlprefix\url{https://doi.org/10.1016/j.ijggc.2017.03.011}.

\bibitem{Gooya2019}
\bibinfo{author}{Gooya, R.} \emph{et~al.}
\newblock \bibinfo{title}{Unstable, super critical {CO}$_2${\textendash}water
  displacement in fine grained porous media under geologic carbon sequestration
  conditions}.
\newblock \emph{\bibinfo{journal}{Scientific Reports}}
  \textbf{\bibinfo{volume}{9}} (\bibinfo{year}{2019}).
\newblock \urlprefix\url{https://doi.org/10.1038/s41598-019-47437-5}.

\bibitem{Kalam2020}
\bibinfo{author}{Kalam, S.} \emph{et~al.}
\newblock \bibinfo{title}{Carbon dioxide sequestration in underground
  formations: review of experimental, modeling, and field studies}.
\newblock \emph{\bibinfo{journal}{Journal of Petroleum Exploration and
  Production Technology}} \textbf{\bibinfo{volume}{11}},
  \bibinfo{pages}{303--325} (\bibinfo{year}{2020}).
\newblock \urlprefix\url{https://doi.org/10.1007/s13202-020-01028-7}.

\bibitem{Hinton2021}
\bibinfo{author}{Hinton, E.~M.} \& \bibinfo{author}{Woods, A.~W.}
\newblock \bibinfo{title}{Capillary trapping in a vertically heterogeneous
  porous layer}.
\newblock \emph{\bibinfo{journal}{Journal of Fluid Mechanics}}
  \textbf{\bibinfo{volume}{910}} (\bibinfo{year}{2021}).
\newblock \urlprefix\url{https://doi.org/10.1017/jfm.2020.972}.

\bibitem{Guo2022}
\bibinfo{author}{Guo, R.} \emph{et~al.}
\newblock \bibinfo{title}{Role of heterogeneous surface wettability on dynamic
  immiscible displacement, capillary pressure, and relative permeability in a
  {CO}$_2$-water-rock system}.
\newblock \emph{\bibinfo{journal}{Advances in Water Resources}}
  \textbf{\bibinfo{volume}{165}}, \bibinfo{pages}{104226}
  (\bibinfo{year}{2022}).
\newblock \urlprefix\url{https://doi.org/10.1016/j.advwatres.2022.104226}.

\bibitem{Alhosani2020}
\bibinfo{author}{Alhosani, A.} \emph{et~al.}
\newblock \bibinfo{title}{Pore-scale mechanisms of {CO}$_2$ storage in
  oilfields}.
\newblock \emph{\bibinfo{journal}{Scientific Reports}}
  \textbf{\bibinfo{volume}{10}} (\bibinfo{year}{2020}).
\newblock \urlprefix\url{https://doi.org/10.1038/s41598-020-65416-z}.

\bibitem{PLUG2007}
\bibinfo{author}{Plug, W.-J.} \& \bibinfo{author}{Bruining, J.}
\newblock \bibinfo{title}{Capillary pressure for the sand–co2–water system
  under various pressure conditions. application to co2 sequestration}.
\newblock \emph{\bibinfo{journal}{Advances in Water Resources}}
  \textbf{\bibinfo{volume}{30}}, \bibinfo{pages}{2339--2353}
  (\bibinfo{year}{2007}).
\newblock
  \urlprefix\url{https://www.sciencedirect.com/science/article/pii/S0309170807000905}.

\bibitem{Iglauer2017}
\bibinfo{author}{Iglauer, S.}
\newblock \bibinfo{title}{{CO}$_2${\textendash}water{\textendash}rock
  wettability: Variability, influencing factors, and implications for {CO}$_2$
  geostorage}.
\newblock \emph{\bibinfo{journal}{Accounts of Chemical Research}}
  \textbf{\bibinfo{volume}{50}}, \bibinfo{pages}{1134--1142}
  (\bibinfo{year}{2017}).
\newblock \urlprefix\url{https://doi.org/10.1021/acs.accounts.6b00602}.

\bibitem{huang2005}
\bibinfo{author}{Huang, H.}, \bibinfo{author}{Meakin, P.} \&
  \bibinfo{author}{Liu, M.}
\newblock \bibinfo{title}{Computer simulation of two-phase immiscible fluid
  motion in unsaturated complex fractures using a volume of fluid method}.
\newblock \emph{\bibinfo{journal}{Water Resources Research}}
  \textbf{\bibinfo{volume}{41}} (\bibinfo{year}{2005}).
\newblock
  \urlprefix\url{https://agupubs.onlinelibrary.wiley.com/doi/abs/10.1029/2005WR004204}.

\bibitem{Pan2004}
\bibinfo{author}{Pan, C.}, \bibinfo{author}{Hilpert, M.} \&
  \bibinfo{author}{Miller, C.~T.}
\newblock \bibinfo{title}{Lattice-boltzmann simulation of two-phase flow in
  porous media}.
\newblock \emph{\bibinfo{journal}{Water Resources Research}}
  \textbf{\bibinfo{volume}{40}} (\bibinfo{year}{2004}).
\newblock
  \urlprefix\url{https://agupubs.onlinelibrary.wiley.com/doi/abs/10.1029/2003WR002120}.

\bibitem{Valvatne2004}
\bibinfo{author}{Valvatne, P.~H.} \& \bibinfo{author}{Blunt, M.~J.}
\newblock \bibinfo{title}{Predictive pore-scale modeling of two-phase flow in
  mixed wet media}.
\newblock \emph{\bibinfo{journal}{Water Resources Research}}
  \textbf{\bibinfo{volume}{40}} (\bibinfo{year}{2004}).
\newblock
  \urlprefix\url{https://agupubs.onlinelibrary.wiley.com/doi/abs/10.1029/2003WR002627}.

\bibitem{Raeini2017}
\bibinfo{author}{Raeini, A.~Q.}, \bibinfo{author}{Bijeljic, B.} \&
  \bibinfo{author}{Blunt, M.~J.}
\newblock \bibinfo{title}{Generalized network modeling: Network extraction as a
  coarse-scale discretization of the void space of porous media}.
\newblock \emph{\bibinfo{journal}{Phys. Rev. E}} \textbf{\bibinfo{volume}{96}},
  \bibinfo{pages}{013312} (\bibinfo{year}{2017}).
\newblock \urlprefix\url{https://link.aps.org/doi/10.1103/PhysRevE.96.013312}.

\bibitem{Neumann2021}
\bibinfo{author}{Neumann, R.~F.} \emph{et~al.}
\newblock \bibinfo{title}{High accuracy capillary network representation in
  digital rock reveals permeability scaling functions}.
\newblock \emph{\bibinfo{journal}{Scientific Reports}}
  \textbf{\bibinfo{volume}{11}} (\bibinfo{year}{2021}).
\newblock \urlprefix\url{https://doi.org/10.1038/s41598-021-90090-0}.

\bibitem{TirapuGHGT}
\bibinfo{author}{Tirapu-Azpiroz, J.} \emph{et~al.}
\newblock \bibinfo{title}{{Cloud-based pore-scale simulator for studying carbon
  dioxide flow in digital rocks}}.
\newblock In \emph{\bibinfo{booktitle}{Proceedings of the 16th Greenhouse Gas
  Control Technologies Conference (GHGT-16)}} (\bibinfo{year}{2022}).
\newblock \urlprefix\url{https://ssrn.com/abstract=4276744}.

\bibitem{lucas2022micro}
\bibinfo{author}{Lucas-Oliveira, E.} \emph{et~al.}
\newblock \bibinfo{title}{Micro-computed tomography of sandstone rocks: Raw,
  filtered and segmented datasets}.
\newblock \emph{\bibinfo{journal}{Data in Brief}} \textbf{\bibinfo{volume}{41}}
  (\bibinfo{year}{2022}).

\bibitem{Esteves2023}
\bibinfo{author}{Esteves~Ferreira, M.} \emph{et~al.}
\newblock \bibinfo{title}{Full scale, microscopically resolved tomographies of
  sandstone and carbonate rocks augmented by experimental porosity and
  permeability values}.
\newblock \emph{\bibinfo{journal}{Scientific Data}}
  \textbf{\bibinfo{volume}{10}} (\bibinfo{year}{2023}).
\newblock \urlprefix\url{https://doi.org/10.1038/s41597-023-02259-z}.

\bibitem{NeumannDataset2020}
\bibinfo{author}{Neumann, R.}, \bibinfo{author}{Andreeta, M.} \&
  \bibinfo{author}{Lucas-Oliveira, E.}
\newblock \bibinfo{title}{11 sandstones: raw, filtered and segmented data}.
\newblock
  \bibinfo{howpublished}{\url{http://www.digitalrocksportal.org/projects/317}}
  (\bibinfo{year}{2020}).

\bibitem{Esteves_Ferreira2023}
\bibinfo{author}{Ferreira, M.~E.} \emph{et~al.}
\newblock \bibinfo{title}{{Full scale, microscopically resolved tomographies of
  sandstone and carbonate rocks augmented by experimental porosity and
  permeability values}}.
\newblock \emph{\bibinfo{journal}{Figshare}}  (\bibinfo{year}{2023}).
\newblock \urlprefix\url{https://doi.org/10.25452/figshare.plus.21375565.v6}.

\bibitem{ZARROUK201913}
\bibinfo{author}{Zarrouk, S.~J.} \& \bibinfo{author}{McLean, K.}
\newblock \bibinfo{title}{Chapter 2 - geothermal systems}.
\newblock In \bibinfo{editor}{Zarrouk, S.~J.} \& \bibinfo{editor}{McLean, K.}
  (eds.) \emph{\bibinfo{booktitle}{Geothermal Well Test Analysis}},
  \bibinfo{pages}{13--38} (\bibinfo{publisher}{Academic Press},
  \bibinfo{year}{2019}).
\newblock
  \urlprefix\url{https://www.sciencedirect.com/science/article/pii/B9780128149461000025}.

\bibitem{Huber2009}
\bibinfo{author}{Huber, M.~L.}, \bibinfo{author}{Perkins, R.~A.},
  \bibinfo{author}{Laesecke, A.} \& \bibinfo{author}{Friend, D.~G.}
\newblock \bibinfo{title}{New international formulation for the viscosity of
  {H}$_2${O}}.
\newblock \emph{\bibinfo{journal}{Journal of Physical and Chemical Reference
  Data}} \textbf{\bibinfo{volume}{38}}, \bibinfo{pages}{101--125}
  (\bibinfo{year}{2009}).
\newblock \urlprefix\url{https://doi.org/10.1063/1.3088050}.

\bibitem{HEIDARYAN2011}
\bibinfo{author}{Heidaryan, E.}, \bibinfo{author}{Hatami, T.},
  \bibinfo{author}{Rahimi, M.} \& \bibinfo{author}{Moghadasi, J.}
\newblock \bibinfo{title}{Viscosity of pure carbon dioxide at supercritical
  region: Measurement and correlation approach}.
\newblock \emph{\bibinfo{journal}{The Journal of Supercritical Fluids}}
  \textbf{\bibinfo{volume}{56}}, \bibinfo{pages}{144--151}
  (\bibinfo{year}{2011}).
\newblock
  \urlprefix\url{https://www.sciencedirect.com/science/article/pii/S0896844610005127}.

\bibitem{Bachu2009}
\bibinfo{author}{Bachu, S.} \& \bibinfo{author}{Bennion, D.~B.}
\newblock \bibinfo{title}{Interfacial tension between {CO}$_2$, freshwater, and
  brine in the range of pressure from (2 to 27) {MPa}, temperature from (20 to
  125)\degree{C}, and water salinity from (0 to 334{\hspace{0.167em}}000)
  mg$\cdotp${L}$^{-1}$}.
\newblock \emph{\bibinfo{journal}{Journal of Chemical {\&} Engineering Data}}
  \textbf{\bibinfo{volume}{54}}, \bibinfo{pages}{765--775}
  (\bibinfo{year}{2009}).
\newblock \urlprefix\url{https://doi.org/10.1021/je800529x}.

\bibitem{Wagner2002}
\bibinfo{author}{Wagner, W.} \& \bibinfo{author}{Pruß, A.}
\newblock \bibinfo{title}{The {IAPWS} formulation 1995 for the thermodynamic
  properties of ordinary water substance for general and scientific use}.
\newblock \emph{\bibinfo{journal}{Journal of Physical and Chemical Reference
  Data}} \textbf{\bibinfo{volume}{31}}, \bibinfo{pages}{387--535}
  (\bibinfo{year}{2002}).
\newblock \urlprefix\url{https://doi.org/10.1063/1.1461829}.

\bibitem{Iglauer2015}
\bibinfo{author}{Iglauer, S.}, \bibinfo{author}{Pentland, C.~H.} \&
  \bibinfo{author}{Busch, A.}
\newblock \bibinfo{title}{Co2 wettability of seal and reservoir rocks and the
  implications for carbon geo-sequestration}.
\newblock \emph{\bibinfo{journal}{Water Resources Research}}
  \textbf{\bibinfo{volume}{51}}, \bibinfo{pages}{729--774}
  (\bibinfo{year}{2015}).
\newblock
  \urlprefix\url{https://agupubs.onlinelibrary.wiley.com/doi/abs/10.1002/2014WR015553}.

\bibitem{ANDRA201333}
\bibinfo{author}{Andrä, H.} \emph{et~al.}
\newblock \bibinfo{title}{Digital rock physics benchmarks—part ii: Computing
  effective properties}.
\newblock \emph{\bibinfo{journal}{Computers {\&} Geosciences}}
  \textbf{\bibinfo{volume}{50}}, \bibinfo{pages}{33--43}
  (\bibinfo{year}{2013}).
\newblock
  \urlprefix\url{https://www.sciencedirect.com/science/article/pii/S0098300412003172}.

\bibitem{Dijkstra59}
\bibinfo{author}{Dijkstra, E.~W.}
\newblock \bibinfo{title}{A note on two problems in connexion with graphs.}
\newblock \emph{\bibinfo{journal}{Numerische Mathematik}}
  \textbf{\bibinfo{volume}{1}}, \bibinfo{pages}{269–271}
  (\bibinfo{year}{1959}).
\newblock \urlprefix\url{https://doi.org/10.1007/BF01386390}.

\bibitem{Raoof2009}
\bibinfo{author}{Raoof, A.} \& \bibinfo{author}{Hassanizadeh, S.~M.}
\newblock \bibinfo{title}{A new method for generating pore-network models of
  porous media}.
\newblock \emph{\bibinfo{journal}{Transport in Porous Media}}
  \textbf{\bibinfo{volume}{81}}, \bibinfo{pages}{391--407}
  (\bibinfo{year}{2009}).
\newblock \urlprefix\url{https://doi.org/10.1007/s11242-009-9412-3}.

\bibitem{ST4SD}
\bibinfo{author}{Johnston, M.~A.}, \bibinfo{author}{Vassiliadis, V.},
  \bibinfo{author}{Pomponio, A.} \& \bibinfo{author}{Pyzer-Knapp, E.}
\newblock \bibinfo{title}{{Simulation Toolkit for Scientific Discovery}}
  (\bibinfo{year}{2022}).
\newblock \urlprefix\url{https://github.com/st4sd/}.

\bibitem{LucasOliveira2022}
\bibinfo{author}{Lucas-Oliveira, E.} \emph{et~al.}
\newblock \bibinfo{title}{Micro-computed tomography of sandstone rocks: Raw,
  filtered and segmented datasets}.
\newblock \emph{\bibinfo{journal}{Data in Brief}}
  \textbf{\bibinfo{volume}{41}}, \bibinfo{pages}{107893}
  (\bibinfo{year}{2022}).
\newblock \urlprefix\url{https://doi.org/10.1016/j.dib.2022.107893}.

\bibitem{Buades2011}
\bibinfo{author}{Buades, A.}, \bibinfo{author}{Coll, B.} \&
  \bibinfo{author}{Morel, J.-M.}
\newblock \bibinfo{title}{Non-local means denoising}.
\newblock \emph{\bibinfo{journal}{Image Processing On Line}}
  \textbf{\bibinfo{volume}{1}}, \bibinfo{pages}{208--212}
  (\bibinfo{year}{2011}).
\newblock \urlprefix\url{https://doi.org/10.5201/ipol.2011.bcm_nlm}.

\bibitem{Darbon2008}
\bibinfo{author}{Darbon, J.}, \bibinfo{author}{Cunha, A.},
  \bibinfo{author}{Chan, T.~F.}, \bibinfo{author}{Osher, S.} \&
  \bibinfo{author}{Jensen, G.~J.}
\newblock \bibinfo{title}{Fast nonlocal filtering applied to electron
  cryomicroscopy}.
\newblock In \emph{\bibinfo{booktitle}{2008 5th {IEEE} International Symposium
  on Biomedical Imaging: From Nano to Macro}} (\bibinfo{publisher}{{IEEE}},
  \bibinfo{year}{2008}).
\newblock \urlprefix\url{https://doi.org/10.1109/isbi.2008.4541250}.

\bibitem{Schindelin2012}
\bibinfo{author}{Schindelin, J.} \emph{et~al.}
\newblock \bibinfo{title}{Fiji: an open-source platform for biological-image
  analysis}.
\newblock \emph{\bibinfo{journal}{Nature Methods}}
  \textbf{\bibinfo{volume}{9}}, \bibinfo{pages}{676--682}
  (\bibinfo{year}{2012}).
\newblock \urlprefix\url{https://doi.org/10.1038/nmeth.2019}.

\bibitem{Immerkr1996}
\bibinfo{author}{Immerk{\ae}r, J.}
\newblock \bibinfo{title}{Fast noise variance estimation}.
\newblock \emph{\bibinfo{journal}{Computer Vision and Image Understanding}}
  \textbf{\bibinfo{volume}{64}}, \bibinfo{pages}{300--302}
  (\bibinfo{year}{1996}).
\newblock \urlprefix\url{https://doi.org/10.1006/cviu.1996.0060}.

\bibitem{Ridler1978}
\bibinfo{author}{Ridler, T.~W.} \& \bibinfo{author}{Calvard, S.}
\newblock \bibinfo{title}{Picture thresholding using an iterative selection
  method}.
\newblock \emph{\bibinfo{journal}{{IEEE} Transactions on Systems, Man, and
  Cybernetics}} \textbf{\bibinfo{volume}{8}}, \bibinfo{pages}{630--632}
  (\bibinfo{year}{1978}).
\newblock \urlprefix\url{https://doi.org/10.1109/tsmc.1978.4310039}.

\bibitem{Liao2001AFA}
\bibinfo{author}{Liao, P.-S.}, \bibinfo{author}{Chen, T.-S.} \&
  \bibinfo{author}{Chung, P.~C.}
\newblock \bibinfo{title}{A fast algorithm for multilevel thresholding}.
\newblock \emph{\bibinfo{journal}{J. Inf. Sci. Eng.}}
  \textbf{\bibinfo{volume}{17}}, \bibinfo{pages}{713--727}
  (\bibinfo{year}{2001}).

\bibitem{Hoshen1997}
\bibinfo{author}{Hoshen, J.}, \bibinfo{author}{Berry, M.~W.} \&
  \bibinfo{author}{Minser, K.~S.}
\newblock \bibinfo{title}{Percolation and cluster structure parameters: The
  enhanced hoshen-kopelman algorithm}.
\newblock \emph{\bibinfo{journal}{Physical Review E}}
  \textbf{\bibinfo{volume}{56}}, \bibinfo{pages}{1455--1460}
  (\bibinfo{year}{1997}).
\newblock \urlprefix\url{https://doi.org/10.1103/physreve.56.1455}.

\bibitem{Niblack1990}
\bibinfo{author}{Niblack, C.~W.}, \bibinfo{author}{Capson, D.~W.} \&
  \bibinfo{author}{Gibbons, P.~B.}
\newblock \bibinfo{title}{{Generating Skeletons and Centerlines from The Medial
  Axis Transform}}.
\newblock In \emph{\bibinfo{booktitle}{Proceedings of 10th International
  Conference on Pattern Recognition}}, vol.~\bibinfo{volume}{I},
  \bibinfo{pages}{881--885} (\bibinfo{year}{1990}).

\bibitem{Telea2003}
\bibinfo{author}{Telea, A.} \& \bibinfo{author}{Vilanova, A.}
\newblock \bibinfo{title}{{A Robust Level-Set Algorithm for Centerline
  Extraction}}.
\newblock In \bibinfo{editor}{Bonneau, G.-P.}, \bibinfo{editor}{Hahmann, S.} \&
  \bibinfo{editor}{Hansen, C.~D.} (eds.) \emph{\bibinfo{booktitle}{Eurographics
  / IEEE VGTC Symposium on Visualization}} (\bibinfo{publisher}{The
  Eurographics Association}, \bibinfo{year}{2003}).
\newblock \urlprefix\url{https://doi.org/10.2312/VisSym/VisSym03/185-194}.

\bibitem{Gostick2016}
\bibinfo{author}{Gostick, J.} \emph{et~al.}
\newblock \bibinfo{title}{Open{PNM}: A pore network modeling package}.
\newblock \emph{\bibinfo{journal}{Computing in Science {\&} Engineering}}
  \textbf{\bibinfo{volume}{18}}, \bibinfo{pages}{60--74}
  (\bibinfo{year}{2016}).
\newblock \urlprefix\url{https://doi.org/10.1109/MCSE.2016.49}.

\end{thebibliography}


\begin{references}
~
\bibliographystyle{naturemag}
\bibliography{Optimizing_carbon_dioxide_trapping_for_geological_storage_arxiv}
\end{references}


\begin{acknowledgements}
We acknowledge discussions with Marcio Carvalho (PUC-Rio) and  project support by Alexandre Pfeifer and Bruno Flach (both IBM Research). Also, we thank Vassilis Vassiliadis, Alessandro Pomponio, and Michael Johnston (all IBM Research) for expert technical assistance.

\end{acknowledgements}



\begin{compet}
The authors declare no Competing Financial or Non-Financial Interests. 
\end{compet}

\begin{additional}
\subsection{Correspondence}
and requests for materials should be addressed to mathiast@br.ibm.com

\newpage

\renewcommand{\thesection}{S\arabic{section}}
\renewcommand{\theequation}{S\arabic{equation}}
\renewcommand{\thesupptable}{S\arabic{supptable}}
\renewcommand{\thesuppfigure}{S\arabic{suppfigure}}

\begin{center}
\Huge{Supplementary Information}    
\end{center}

\section{Simulation scenarios and parameters}\label{sec:SimulationParameters} 

Four temperature scenarios were investigated in this work, representing increasingly deeper injection points in reservoirs and resulting in corresponding changes to viscosity and density of each of the fluids considered, and to the fluid-fluid interfacial tension. 
Appropriate values for those parameters relevant to our simulations were extracted from the literature \cite{Bachu2009}. 
Whenever not available, we calculated supercritical {CO}$_2$ viscosity\cite{HEIDARYAN2011} and water viscosity\cite{Huber2009,Wagner2002} based on equations of state obtained from experimental data for the temperatures of interest.
Table \ref{ParameterTable} collects the parameters values used  in our simulations. 

\graphicspath{{./figures/}}
\begin{supptable}[ht]
  \includegraphics[width=\linewidth]{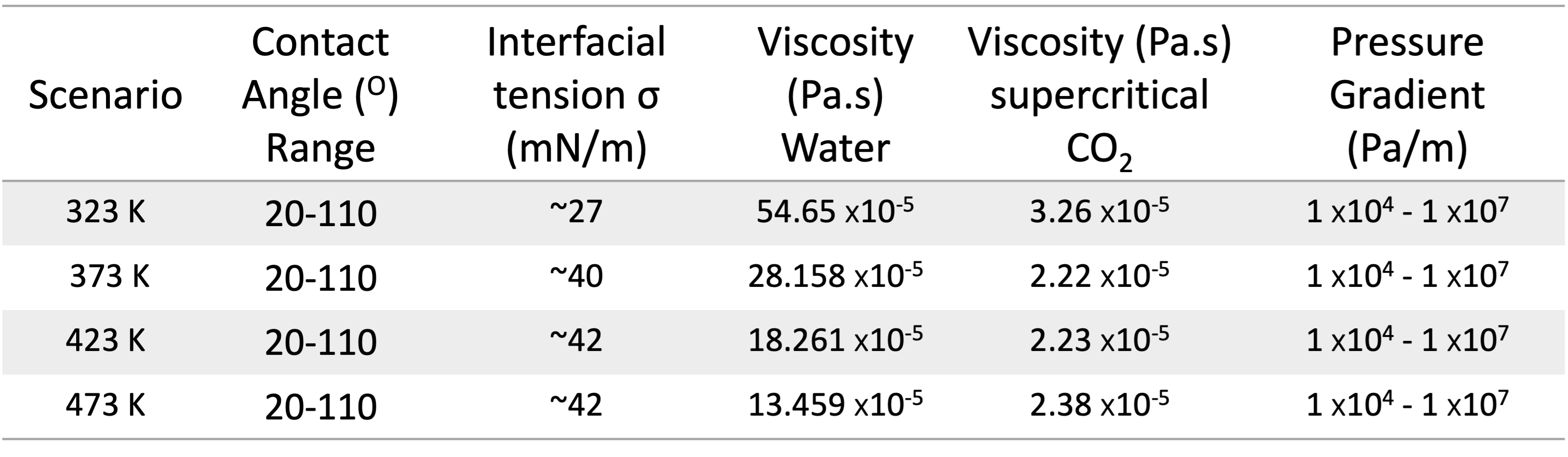}
  \caption{\textbf{Simulation scenarios.} Set of parameters used in the two-phase simulations, for each temperature scenario of interest.}
  \label{ParameterTable}
\end{supptable}

\section{Digital Rock Image processing}\label{sec:SuppDigitalRock} 

We used a large dataset of three-dimensional images extracted from X-ray computed micro-tomography ($\mu$CT) scans of sandstone and carbonate rock samples at a resolution of $2.25\,\mu\text{m/voxel}$.\cite{Neumann2021,Esteves2023} The $\mu$CT system scans cylindrical rock plugs, as seen in Fig. \ref{DigitalRock}a. A 26 mm-long and 10 mm-wide rock plug suitable for our system (Skyscan 1272, Bruker), was imaged as two-dimensional projections of the pore space. 
The three-dimensional digital rock image is reconstructed out of these two-dimensional projections using the built-in Bruker software (NRecon, version 1.7.0.4, with the Reconstruction engine InstaRecon, version 2.0.2.6). 
The reconstructed volume of the fully digitized rock sample obtained from the $\mu$CT measurements is usually cropped into smaller volumes that are more computationally manageable, while retaining statistically significant rock properties. For a set of sandstone rock samples scanned at a resolution of $2.25\,\mu\text{m/voxel}$, it was observed that a Representative Elementary Volume (REV) of about $1000^3$ voxels was sufficient to yield accurate rock property predictions like porosity and permeability\cite{Neumann2021}.
Additional image processing methods are applied to the digitized rock following the procedures described in references\cite{Neumann2021,LucasOliveira2022} to equalize contrast differences due to mineralogical compositions across images.  To fully eliminate image and measurement artifacts resulting in very bright spots in the image, all voxels above a 99.8$\%$ threshold in the grayscale cumulative histogram have been removed and the remaining grayscale levels have been remapped to span the full [0, 255] scale.
To reduce noise, a 3D non-local means filter\cite{Buades2011,Darbon2008} was applied, which is available in Fiji\cite{Schindelin2012}, using a smoothing factor of 1 and automatically estimated sigma\cite{Immerkr1996} parameters.

Finally, several thresholding algorithms have been implemented to segment the grayscale images and separate the pore space (darker) from the rock matrix (lighter), depending on the rock type. For instance, using a threshold level calculated by the IsoData method\cite{Ridler1978} was sufficient to accurately segment the sandstone grayscale images into solid (white) and void (black) space leading to binary images\cite{Neumann2021}, but a 3-level multilevel Otsu method\cite{Liao2001AFA} was needed to properly group sub-porous regions, with little expected flow, with the mineral matrix for carbonate samples\cite{Esteves2023}. Fig. \ref{DigitalRock}b displays an example of this segmentation process as applied to a $225\,\mu\text{m} \times 225\,\mu\text{m}$ cross-section of a carbonate rock tomography, where the grayscale image on the left is segmented into the binary image on the right. The Enhanced Hoshen-Kopelman algorithm\cite{Hoshen1997} is used to locate all pore clusters, determine the pore volume fraction, and eliminate the pore clusters that are not connected to the percolating pore network. The connected pore space is then taken as the true measure of porosity.

\begin{suppfigure}[ht]
  \includegraphics[width=\linewidth]{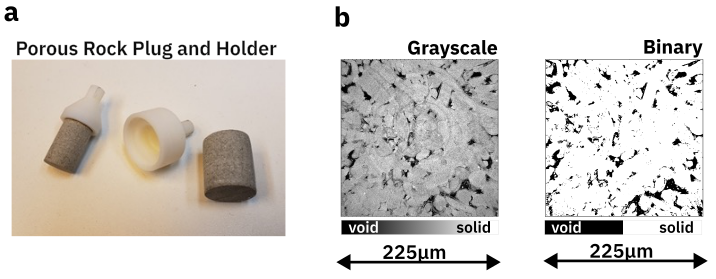}
  \caption{\textbf{Digital rock image processing.} a) Examples of porous rock plugs and their plastic sample holders. b) Tomography cross section with dimensions $225\,\mu\text{m} \times 225\,\mu\text{m}$ in grayscale (left) and its corresponding binarized image (right), scanned at a resolution of $2.25\,\mu\text{m/voxel}$.}
  \label{DigitalRock}
\end{suppfigure}

\section{Capillary Network extraction and centerlines representation}\label{sec:CNM} 

The CNM representation is based on the Centerline extraction algorithm\cite{Neumann2021}. A centerline is a thin, one-dimensional object that captures a 3D object’s main symmetry axes, summarizing its main shape into a set of curves\cite{Niblack1990,Telea2003}.
Starting from the 3D binary image, our network extraction algorithm transforms the pore space into voxel-wide lines at the center of the pore channels, finding the most central paths from inlet pores to outlet pores through a centrality-based cost function inspired by the Dijkstra’s Minimum Path algorithm \cite{Dijkstra59}.
The resulting graph is interpreted as a cylindrical capillary (a short cylinder), one voxel long, but whose diameter is the distance to the closest solid voxel.
The full algorithm description can be found in Neumann \textit{et al.}\cite{Neumann2021}.

The digital rock image in Fig. \ref{Centerlines}a has dimension of $(225\,\mu\text{m})^3$ and was digitized in 8-bit grayscale using one million (100$^3$) voxels, each representing a physical region of $(2.25\,\mu\text{m})^3$. After segmentation, about 26$\%$ of the voxels in the resulting binary image of Fig. \ref{Centerlines}b, are identified as pore space, 64.5$\%$ of which touch the rock surface. After undergoing the extraction procedure, the capillary network model of the sample provides an accurate representation of the rock porous geometry as seen in Fig. \ref{Centerlines}c. This CNM representation contains 4069 capillaries with a color-coded diameter scale. The capillary network of the Berea sandstone sample analyzed in this paper, of size equalto $1000^3$ voxels, displays a porosity of 33$\%$ and results in a CNM with $2.7 \times 10^6$ capillaries.

\begin{suppfigure}[ht]
  \includegraphics[width=\linewidth]{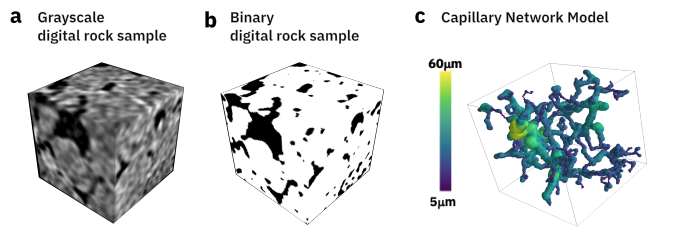}
  \caption{\textbf{Capillary Network extraction}. 3D grayscale digital rock image (a) and 3D binary image (b) of a Berea sandstone sample with 100$^3$ voxels. Capillary network representation (c) of the sample mapping the rock pore geometry with a color-coded diameter}
  \label{Centerlines}
\end{suppfigure}

\section{Single-phase flow simulation}\label{sec:SinglePhaseFlow}

The fined-grained capillary network representation of the rock's pore space described in section \ref{sec:CNM} was employed to simulate both single and two-phase fluid flow with a high level of geometrical accuracy. 
We assume laminar flow and apply equations relating pressure and flow rate within each capillary, followed by conservation of mass at each node, to build a large system of coupled equations in sparse matrix form. 
The Hagen-Poiseuille equation \ref{equation:Poiseuille} is applied when simulating single-phase stationary flow

\begin{equation}\label{equation:Poiseuille}
Q_j=\frac{\pi R_j^4}{8 \mu L_j}  \Delta P_j ,  \quad \quad  \textrm{for every capillary } j
\end{equation}

\noindent
where $R_j$ and $L_j$ denote, respectively, the radius and length of capillary $j$, $\mu$ is the viscosity of the fluid, and $Q_j$ and $\Delta P_j$  represent the flow rate and pressure difference across each capillary $j$. Combining Eq. \ref{equation:Poiseuille} with the conservation of mass at each network node, Eq. \ref{equation:conservation}, 

\begin{equation}\label{equation:conservation}
\sum_j Q_{i,j} = 0,    \quad \quad   \textrm{for every node } i,
\end{equation}

\noindent
results in a large but rather sparse system of linear equations to solve. The system solution contains local properties such as the distribution of pressure and flow rate at each point in the network, as well as global flow properties like permeability\cite{Neumann2021}. 

Fig. \ref{PressureFields} displays the results of single-phase, pressure-driven flow simulations using the capillary network representation of a $1000^3$ voxel sandstone rock sample as input geometry. A pressure gradient of 10 kPa/m between opposite sides along one axis is applied to drive the flow of a simple fluid with density 1000 kg/m$^3$ and dynamic viscosity 1.002 mPa.s, representing water at 1 atm, through the capillary network. Fig \ref{PressureFields} is showing the induced pressure field inside the 3D sandstone sample of dimensions $2.25\,\text{mm} \times 2.25\,\text{mm} \times 2.25\,\text{mm}$. Due to the sparse nature of all matrices involved, single-phase flow simulations are very efficient and can be simulated within minutes, depending on its connected pore structure and the direction of the flow. The absolute permeability for this REV was computed to be 105, 92 and 54 mD along the X-, Y- and Z-axis, respectively.

\begin{suppfigure}[ht]
  \includegraphics[width=\linewidth]{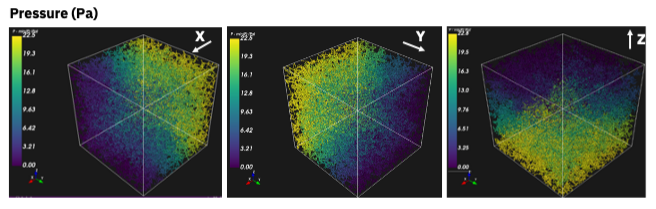}
  \caption{\textbf{Single-phase fluid flow simulation}. 
  Simulated pressure fields inside a capillary network induced by single-phase flow of a a simple fluid representing water at 1 atm . The flow was driven by an external 10 kPa/m pressure gradient imposed across the (a) X-axis (b) Y-axis and (c) Z-axis.}
  \label{PressureFields}
\end{suppfigure}

\section{Two-phase flow simulations}\label{sec:TwoPhaseFlow}

Two-phase flow simulations track the displacement in time of the fluid interface within each capillary of the CNM. We restrict the modeling to incompressible fluids under laminar, one-dimensional flow along the length of each capillary. Each fluid-fluid interface is assumed perpendicular to the flow direction, \textit{i.e.} piston-like displacement. The pressure difference between the ends of each capillary is expressed in Eq. \ref{equation:twophase} as the sum of various physical effects, some depending on the position $x(t)$ of the fluid interface,

\begin{equation}\label{equation:twophase}
{\Delta P}_j\ = \frac{8}{R_j^2}\left[ \mu_1 x_j + \mu_2 \left(L_j - x_j \right) \right] {\dot{x}}_j\ -\frac{2\sigma}{R_j} \cos{\theta} \quad \quad \textrm{for each capillary}\,j ,
\end{equation}

\noindent where ${\Delta P}_j$ is the pressure difference across capillary $j$, with length $L_j$ and radius $R_j$. $\mu_i$ represents the viscosity of fluid $i$ and $\sigma$ the interfacial tension between fluids, while $\theta$ describes the contact angle formed by both fluids and the matrix surface, defined with reference to the injected fluid as illustrated in Table \ref{ContributionsParameterTable}. Finally, $x_j$ and ${\dot{x}}_j$ denote the position of the effective interface and its first derivative, respectively. The first term on the right-hand side refers to viscous forces, and second term refers to the capillary pressure contribution from the interfaces. This contribution cancels out in capillaries with an even number of (alternating) interfaces, and reverts to something similar to Eq. \ref{equation:Poiseuille}. Any odd number of interfaces can be represented by a single ``effective'' interface whose position is calculated as to preserve the relative saturation of each fluid.

\begin{supptable}[!htb]
  \includegraphics[width=\linewidth]{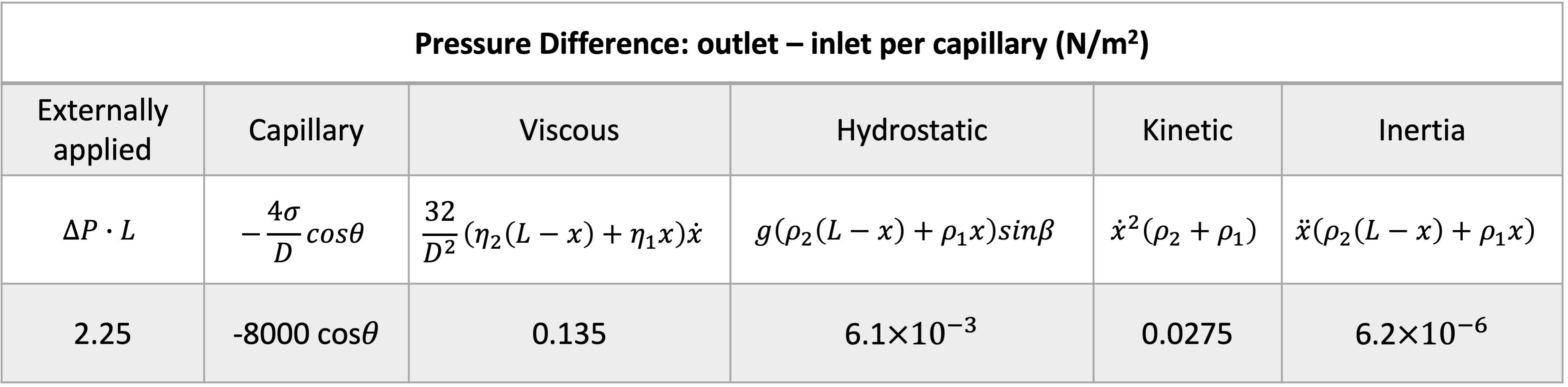}
  \caption{\textbf{Contributions to the pressure gradient per capillary}. Contributions to the pressure difference at both ends of each capillary on a two-phase flow simulation based on typical parameter values.}
  \label{ContributionsTable}
\end{supptable}

In addition to the externally applied pressure gradient, viscous forces and capillary pressure, the physical equations of two-phase flow in capillaries may also include hydrostatic, kinetic, and inertia contributions. Table \ref{ContributionsTable} summarizes the expressions and estimated values of these physical effects, based on the assumptions in Table \ref{ContributionsParameterTable}. The average capillary length (\text{i.e.}, one voxel) and diameter are representative of the distribution shown in Fig. \ref{Fig1_Methodology}b. Under these assumptions, the contributions from hydrostatic, kinetic and inertial forces appear at least one order of magnitude smaller than the least significant of the other three contributions, hence the choce of terms kept in Eq. \ref{equation:twophase}. Together with mass conservation at the nodes, Eq. \ref{equation:twophase} forms a system of differential-algebraic equations (DAE) representing the interface dynamics over time.

\begin{supptable}[!htb]
  \includegraphics[width=\linewidth]{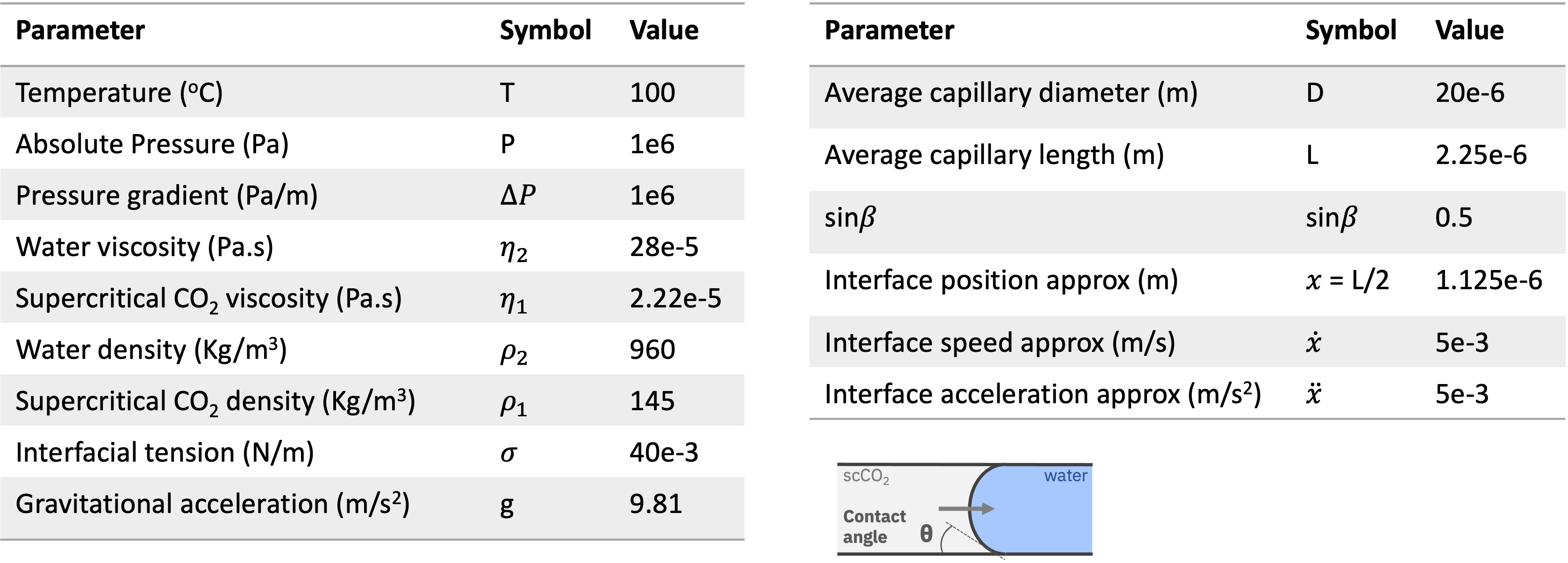}
  \caption{\textbf{Representative parameter values.} Parameters used to estimate the contributions to the pressure difference on a two-phase simulation representative of our study.}
  \label{ContributionsParameterTable}
\end{supptable}

\begin{suppfigure}[!htb]
 \includegraphics[width=\linewidth]{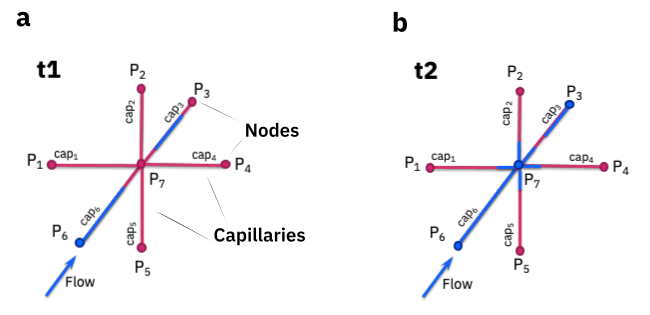}
 \caption{\textbf{Fluid interface evolution in capillary network.} a) Time step where fluid interface progresses through the capillary before reaching a node, and b) time step after fluid interface in \textit{cap$_6$} reaches the central node and transitions to all connected capillaries.}
 \label{InterfacesEvolution}
\end{suppfigure}

Tracking of the fluid interfaces across the network of capillaries proceeds in alternating sequences of \textit{free evolution} and \textit{jumps}. During free evolution, the effective interfaces move along their respective capillaries, but without leaving them, via integration of the DAE system. Fig. \ref{InterfacesEvolution}a displays a simple network of 6 capillaries, represented by \textit{cap$_i$,} where $i=1..6$, sharing a single central node labeled with the pressure \textit{P$_7$} at that point. This example illustrates the free evolution time interval until the fluid interface in \textit{cap$_6$} reaches the central node. Jump events occur when an interface reaches a node and leaves its current capillary to enter one or more neighbouring capillaries. In this step, free evolution pauses, the interfaces are redistributed throughout the network, and the system of DAE is rewritten to account for the changes in interface locations. As an example, in Fig. \ref{InterfacesEvolution}b, an interface that reached the end of one capillary (\textit{cap$_6$}) is removed, and new interfaces are created at the entrances of the connected capillaries. After this rearrangement, free evolution resumes.
Depending on the local pressure state, some capillaries may become plugged, that is, the pressure conditions may not favor flow and the interface becomes frozen at the location of nearest node. It is possible that events such as the merging of fluids, reversal of flow, or changes in the pressure conditions at a later time may start favoring flow again and lead to the capillary becoming unplugged. All possibilities are handled carefully during the redistribution of interfaces among the capillaries.

A more realistic example of the time evolution of a two-phase simulation is displayed in Fig. \ref{TwoPhase_simulations}a and Fig. \ref{TwoPhase_simulations}b, representing the injection of supercritical CO$_2$ over time as it pushes the resident water in a small portion of a Berea sandstone rock modeled as a network of connected capillaries. From the knowledge of the position of the fluid interface in all capillaries, we can compute the saturation as a function of time as displayed in Fig. \ref{TwoPhase_simulations}c. Saturation of the rock sample by the injected fluid an important property in the study of CO$_2$ infiltration into the porous rock. Plotting the saturation as a function of the volume of CO$_2$ injected, as in Fig. \ref{TwoPhase_simulations}d, provides a measure of the injection efficiency. 

\begin{suppfigure}[!htb]
 \includegraphics[width=\linewidth]{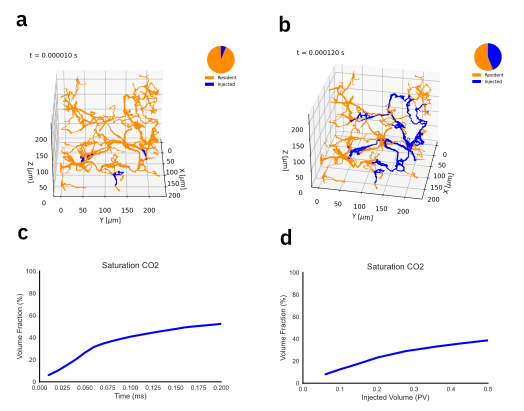}
 \caption{\textbf{Two-phase fluid flow simulation}. Dynamics of two-phase fluid flow simulation at a) time $t_1$ and b) time $t_2$. Measure of scCO2 saturation into the porous rock as c) a function of time and d) a function of injected volume.}
 \label{TwoPhase_simulations}
\end{suppfigure}

\section{Simplified capillary network representation}\label{sec:SimplifiedCNM}

Unlike single-phase (stationary) flow that can be simulated within minutes, even on the network representation of a REV-sized rock sample with millions of nodes and links, dynamic simulations rapidly become unfeasible even on small sample sizes and large computing resources due to their time-dependent nature.  
To overcome this limitation, we employ sets of smaller capillary networks that remain representative of the original capillary network, to run two-phase simulations and extract flow properties. Previous works have demonstrated the importance of the geometric properties of porous media, in particular, the distributions of sizes and shapes of pores and throats, and also topology parameters such as connectivity and coordination number distribution\cite{Raoof2009}. Thus, in order to produce realistic predictions, our smaller and simplified capillary networks are required to accurately match the morphology of the original rock sample. 

To create the simplified capillary network, we have developed a custom Python script based on the OpenPNM\cite{Gostick2016} framework capable of generating arbitrary 3D capillary networks in \textit{regular} (cubic) and \textit{random} configurations.  
A \textit{regular} capillary network comprises a 3D cubic structure delimited by the sample size dimensions $L_X$, $L_Y$ and $L_Z$, where capillaries are created parallel to the axes on a regular mesh, intersecting at regular intervals.  
The intersection between capillaries, referred to as a node, defines its coordination number. One can create different cubic network configurations with coordination numbers from 6 to 26 by connecting faces, edges, corners, and their combination.
In the \textit{random} capillary network, a set of points (or nodes) are randomly distributed in 3D space delimited by the sample dimensions $L_X$, $L_Y$ and $L_Z$. 
In both regular and random capillary network cases, the connections between nodes, when present, become capillaries.

Our simplified capillary networks are built by iterating over the following steps until the porosity of the synthesized network is within a predefined margin from that of the original. In a first step, the coordination number of each node is assigned by choosing from the probability distribution of the original capillary network (see Fig. \ref{SimplifiedCNMcubic}a and Fig. \ref{SimplifiedCNMrandom}a for examples of coordination number probability distribution in a regular and a random capillary network, respectively, overlaid with the distribution of the original sample). Then, capillaries connected to a node are deleted if needed to match the assigned coordination number. In a third step, capillary diameters are assigned by randomly choosing from the diameter probability distribution of the original capillary network (see Fig. \ref{SimplifiedCNMcubic}b and Fig. \ref{SimplifiedCNMrandom}b for examples of capillary diameter probability distribution in a regular and a random capillary network, respectively, overlaid with the distribution of the original sample). 
The final step in each iteration involves calculating the porosity of the simplified network by dividing the capillary volume by the sample volume and comparing it to the porosity of the original rock sample. This algorithm guarantees that the simplified network preserves the following properties: (i) porosity; (ii) capillary diameter distribution; and (iii) node coordination number distribution. Examples of 2D simplified capillary networks in regular and random configurations can be seen in Fig. \ref{SimplifiedCNMcubic}c and Fig. \ref{SimplifiedCNMrandom}c, respectively.  Examples of 3D simplified capillary networks in regular and random configurations can be seen in Fig. \ref{SimplifiedCNMcubic}d and Fig. \ref{SimplifiedCNMrandom}d, respectively.

\begin{suppfigure}[!htb]
  \includegraphics[width=\linewidth]{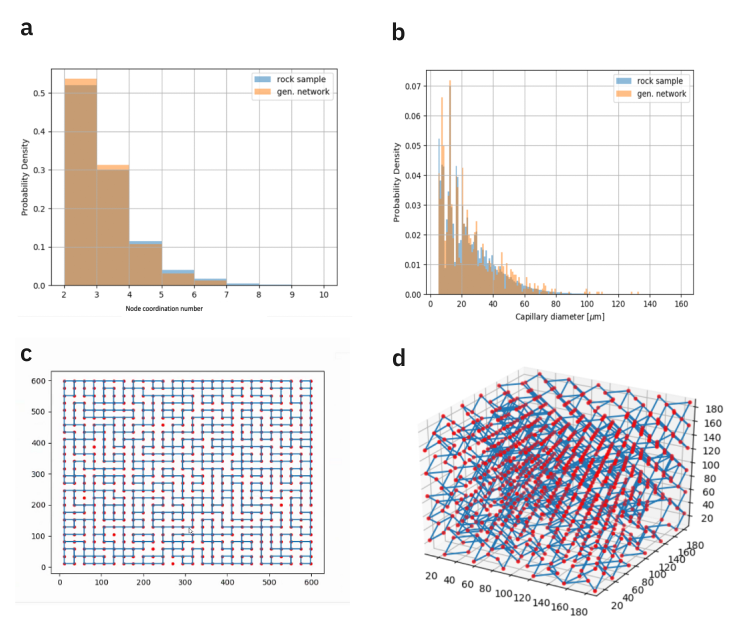}
  \caption{\textbf{Regular simplified capillary network.} a) Node coordination number probability distributions and  b) capillary diameter probability distribution of the synthesized network and the network of the original rock sample. c) 2D regular simplified capillary network with the capillaries represented in blue lines and the nodes in red dots. d) 3D regular simplified capillary network.}
  \label{SimplifiedCNMcubic}
\end{suppfigure}

\begin{suppfigure}[!htb]
  \includegraphics[width=\linewidth]{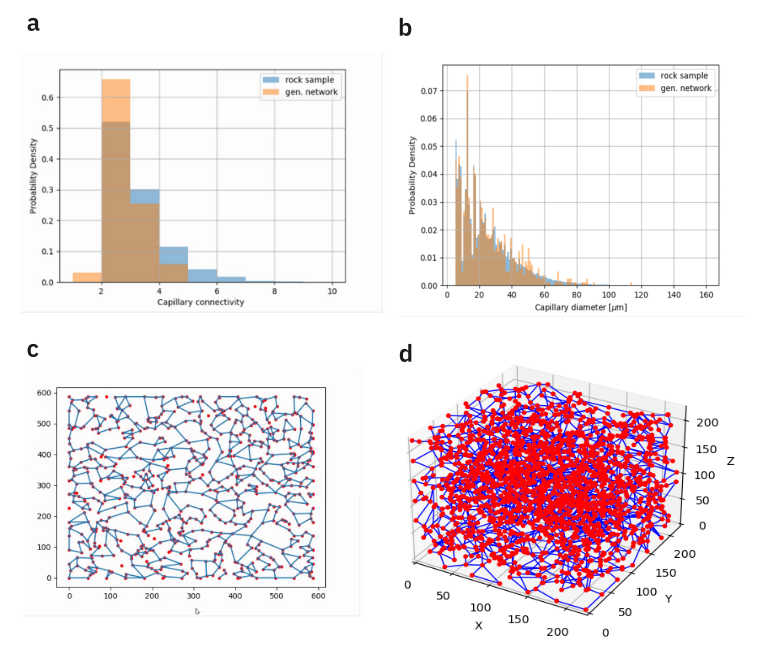}
  \caption{\textbf{Random simplified capillary network.} a) Node coordination number probability distributions and  b) capillary diameter probability distribution of the synthesized network and the network of the original rock sample. c) 2D random simplified capillary network with the capillaries represented in blue lines and the nodes in red dots. d) 3D random simplified capillary network.}
  \label{SimplifiedCNMrandom}
\end{suppfigure}

Single-phase permeability calculations on $\sim$50 different network configurations optimized to represent the original Berea sandstone sample are shown in Fig. \ref{PermvsCNMsize}. The flow was imposed by applying an external 10 kPa/m pressure gradient along each axis. From these results, we conclude that a network size of 1500 capillaries, about $0.5\%$ of the total number of capillaries in the original sample, represents a good trade-off between accuracy and computational cost, with an average permeability within $\pm 3 \sigma$ of the original. We observe that larger representations with more capillaries do improve precision but not the accuracy of the average estimate.

\begin{suppfigure}[ht!]
  \includegraphics[width=\linewidth]{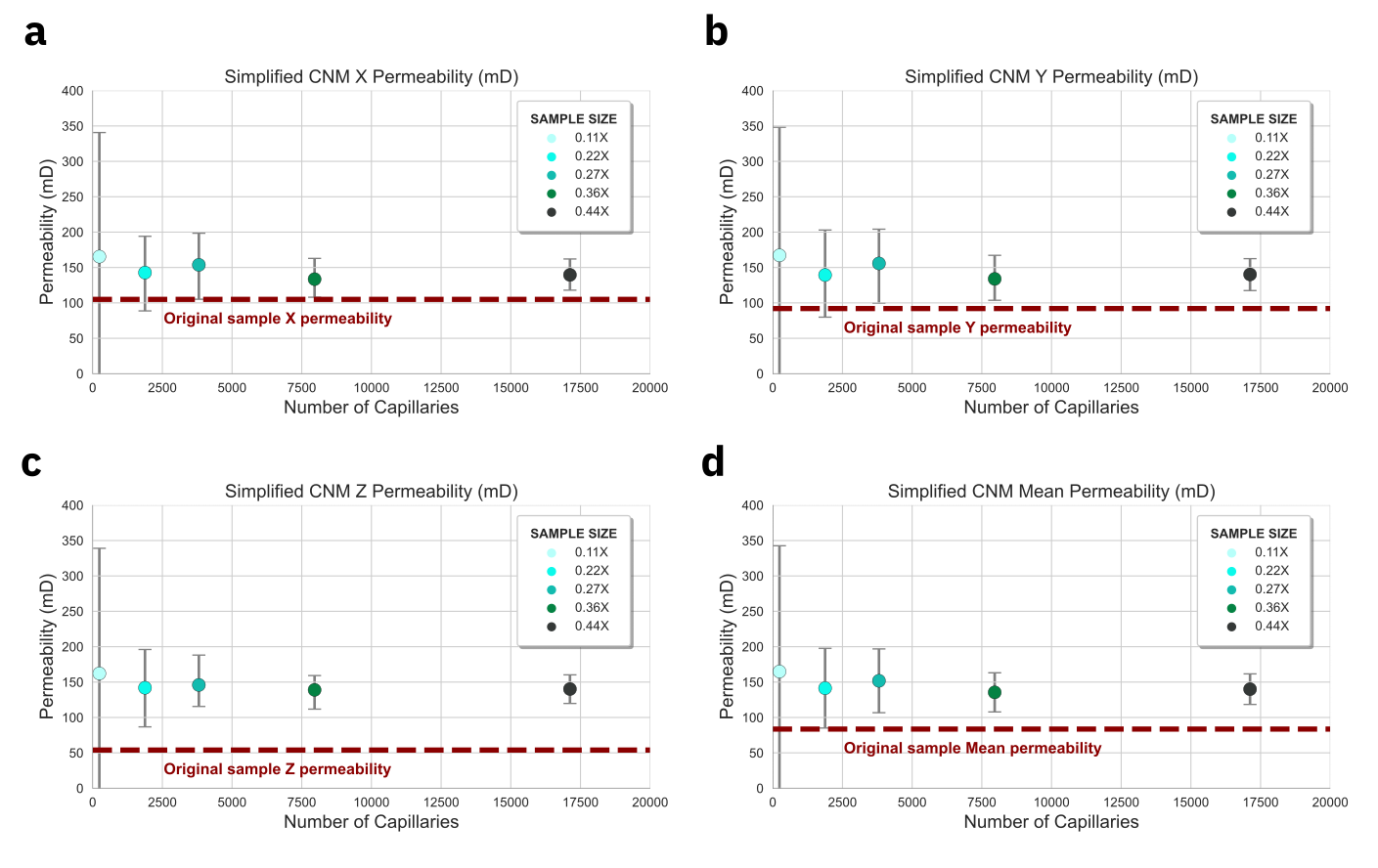}
  \caption{\textbf{Permeability vs ssCN Size.} Mean permeability over the ensenble of $\sim 50$ simplified random network configurations along the (a) X-axis, (b) Y-axis, (c) Z-axis and (d) the mean across all three axes, as a function of the number of capillaries per network.}
  \label{PermvsCNMsize}
\end{suppfigure}

\section{Pressure Distribution Within Capillary Network}\label{sec:CNMPressures}

The resulting distribution of pressures after simulating the injection of scCO$_2$ into one of the simplified capillary networks is plotted in Fig. \ref{SuppPressuresCombined} under varying conditions of externally applied pressure gradient and fluid interface contact angle. This pressure distribution is the result of the interplay between the driving pressure gradient and the internal viscous and capillary forces in each capillary as described in section \ref{sec:TwoPhaseFlow}. 
For the range of capillary diameters shown in Fig.\ref{Fig1_Methodology}b, capillary pressures within the network are estimated to be around 4 kPa for contact angles between 20$^\circ$ and 80$^\circ$, as per the second term of right hand side of Eq. \ref{equation:twophase}. We observe in Fig. \ref{SuppPressuresCombined}a that a driving pressure gradient around $1\times10^4$ Pa/m is not sufficient to overcome the capillary pressures within the network for a contact angle of $20^\circ$. As a consequence, the fluid flow in this low pressure gradient regime is mostly driven by capillary pressure which tends to produce low CO$_2$ saturation. 

For stronger externally applied pressure gradients, we observe a significant change in the pressure distribution of Fig. \ref{SuppPressuresCombined}a and \ref{SuppPressuresCombined}b towards higher pressures for both 20$^\circ$ and 90$^\circ$ contact angles. 
As the pressures induced by the external driving force are now sufficient to overcome the opposing viscous and capillary forces, even under conditions of low contact angles, it is thus able to unplug many capillaries and reach higher saturation values. For the flow condition depicted in Fig. \ref{SuppPressuresCombined}a, namely, $1\times10^7$ Pa/m pressure gradient and $20^\circ$ contact angle, we obtained saturation values of 20\% to 40\%. But, more strikingly, for the condition represented in Fig. \ref{SuppPressuresCombined}b, \textit{i.e.}, $1\times10^7$ Pa/m pressure gradient and $90^\circ$ contact angle, we obtained saturation values reaching 80\%.

\begin{suppfigure}[ht!]
  \includegraphics[width=\linewidth]{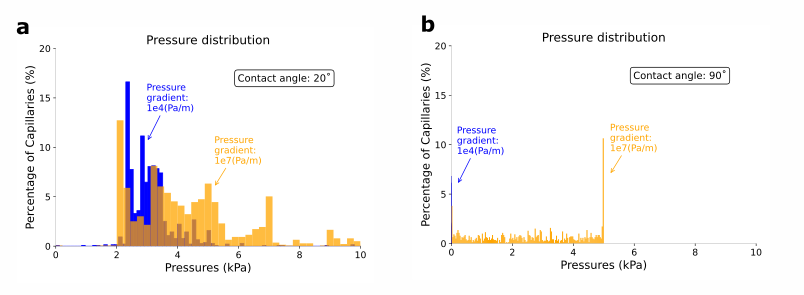}
  \caption{\textbf{Pressure distributions within the CNM} Distribution of resulting pressures at the ends of capillaries for (a) contact angle $20^\circ$, and pressure gradients $1\times 10^4$ Pa/m and $1\times 10^7$ Pa/m; (b) contact angle $90^\circ$ and pressure gradients $1\times 10^4$ Pa/m and $1\times 10^7$ Pa/m.}
  \label{SuppPressuresCombined}
\end{suppfigure}

\section{Saturation vs Injected Volume}\label{sec:Saturation}

Fig. \ref{SuppSatvsInjVol}a shows an example of supercritical CO$_2$ saturation as a function of time in the sandstone rock sample under study at a temperature of 473 K and an external pressure gradient of $5\times 10^6$ Pa/m for a range of contact angles. The curves and shaded areas in the plot represent the mean and standard deviation of the simulations performed in an ensemble of $\sim 50$ ssCN, along the X, Y and Z axis. Alternatively we can display the saturation as a function of the scCO$_2$ injected volume (in units of ``pore volume''), as shown in Fig. \ref{SuppSatvsInjVol}b for the same simulation results. 
To reduce the influence of backward flow, the injected volume in this context is computed from the sum of the scCO$_2$ saturation at each time step plus the volume of scCO$_2$ ejected from the outlet capillaries, calculated by integrating over time the product of the outlet capillaries cross sectional area and the local flow speeds, normalized by the total pore volume occupied by all capillaries in the network. 

Fig. \ref{SuppSatvsInjVol}b shows that for most contact angles, the saturation of scCO$_2$ reaches a plateau, indicating that fluid is not being retained within the pore space.
As a measure of the injection efficiency, we define the variable \textit{Weighted Saturation} ($wS$) as the measured saturation (\textit{S}) scaled by the ratio of saturation to injected volume (\textit{IV}), that is, $wS = S\frac{S}{IV}$, with units of pore volume of injected CO$_2$.
As the saturation plateaus, the value of the $wS$ peaks and any additional injection does not result in further storage by capillary trapping. This variable thus emphasizes the effect that the CO$_2$ injected volume has in the maximum achievable saturation, as observed in Fig. \ref{SuppSatvsInjVol}c (for a range of pressure gradients, with fixed $85^\circ$ contact angle) and Fig. \ref{SuppSatvsInjVol}d (for a range of contact angles, with fixed $5\times10^6$ Pa/m pressure gradient).

\begin{suppfigure}[ht!]
  \includegraphics[width=\linewidth]{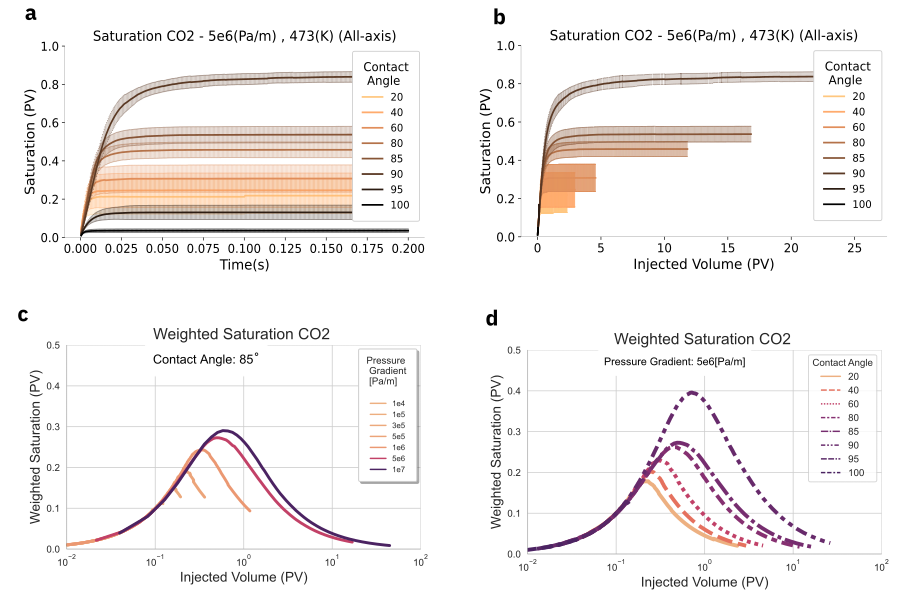}
  \caption{ \textbf{Weighted Saturation \textit{vs.} Injected Volume}. Outcome of two-phase simulations measured as the mean and standard deviation of the results from an ensemble of 50 ssCN, assuming supercritical CO$_2$ as injected fluid, water as resident fluid, a temperature of 473 K and an applied pressure of $1\times10^6$ Pa/m. (a) Saturation of supercritical CO$_2$ as a function of time for various values of contact angle. (b) Saturation as a function of injected volume across various values of contact angle. (c) Weighted saturation as a function of injected volume for a fixed contact angle of $85^\circ$ and a range of applied pressure gradients, and (d) weighted saturation as a function of injected volume for a fixed applied pressure of 5$\times10^6$ Pa/m and the range of contact angles.}
  \label{SuppSatvsInjVol}
\end{suppfigure}

\section{Simulation Toolkit for Scientific Discovery (ST4SD)}\label{sec:ST4SD}

In our study we scanned through 4 temperature scenarios, studied 8 fluid-interface contact angles per scenario and no less than 8 different driving pressure gradient cases per angle, totaling 256 different injection conditions to be simulated. Per injection condition, 150 flow simulations were executed applying driving pressure along all three axes on each of the 50 simplified capillary networks in the ensemble, requiring proper parsing and aggregation of nearly 40,000 simulations.  
To manage the large parameter space, we leveraged high-throughput automated simulation workflows. In particular, we employed the Simulation Toolkit for Scientific Discovery (ST4SD)~\cite{ST4SD} to automate the execution of long simulation campaigns with several chained steps. The use of such workflow scheduler ensures the reproducibility of our results and enable efficiency gains by optimising the use of computing resources.

\begin{suppfigure}[ht]
 \includegraphics[width=\linewidth]{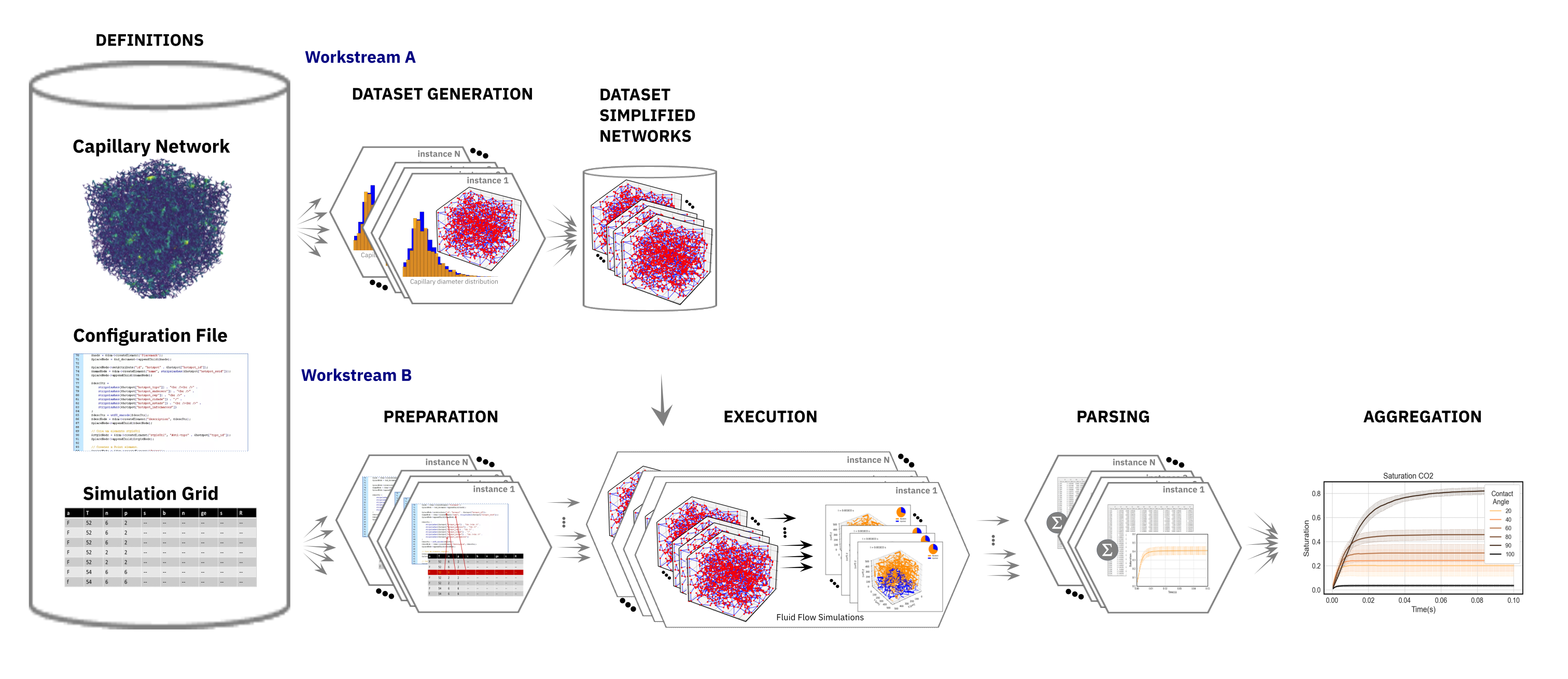}
 \caption{\textbf{Schematic ST4SD workflow.} Automated computational methodology showing, from left to right, experiment definition, preparation of inputs, simulation execution, output parsing, and aggregation of results}
 \label{ST4SD}
\end{suppfigure}

Fig. \ref{Fig1_Methodology} illustrates the conceptual workflow and Fig. \ref{ST4SD} shows the sequence of steps executed in an ST4SD experiment. A CNM representation of a rock sample is used as input to the ST4SD routine. 
This capillary network model is, however, too detailed to solve numerically in a two-phase flow scenario, so we generate tens of simplified capillary networks that meaningfully represent the properties of the original network (see Supplementary Section \ref{sec:SimplifiedCNM}). Each simplified capillary network is then used as  input to independent flow simulations that will estimate relevant physical properties in each representative network. Finally, the individual results from each simplified network are aggregated and combined into a single estimate that applies to the reference network. 

Fig. \ref{ST4SD} shows two connected workstreams. Workstream \textit{A} refers to the process of generating a large ensemble of simplified capillary networks from that of a high-resolution digital rock sample. The Dataset Generation step follows the methodology described in Supplementary Section \ref{sec:SimplifiedCNM}. Taking the original rock sample CNM as input, the workflow launches in parallel many processes to generate simplified capillary networks. In each parallel process, the algorithm alternates between molding a (initially random) set of connected capillaries into matching the morphological properties of the reference rock, and running single-phase flow simulations to assess the permeability until a convergence criterion is reached. The outcome of this workstream is an ensemble of simplified capillary networks whose morphological properties mimic those of the original network.

Workstream \textit{B} refers to the simulated injection of scCO$_2$ on an ensemble of simplified capillary networks and extracting relevant properties from the aggregate of the results. A simulation parameter grid containing all the values to be executed is used as input to the \textit{preparation} step of workstream \textit{B}. Per instance of this simulation grid, the values of parameters such as applied pressure, temperature or contact angle are inserted into the configuration files of the simplified capillary networks produced in workstream \textit{A}. In the \textit{execution} step, two-phase fluid flow simulations are executed for the ensemble of networks and the interfaces are tracked within the capillaries to extract saturation values as a function of time. During \textit{parsing} the saturation of all networks are averaged as a function of time and injected volume, and passed to the \textit{aggregation} step where they are saved together with the results from other instances.

\section{Injection Safety and Efficiency}\label{sec:MaxWSvsTemp}

Fig. \ref{Fig3_InjectionConditions} of the main manuscript explores the efficiency and security of the scCO$_2$ drainage process in deep reservoirs at a fixed 473 K temperature. 
Fig. \ref{MaxSatvsinjVolvsTemp} explores the safety of the process by plotting the value of saturation at 90$\%$ of the maximum \textit{vs.} the injected volume required to reach that point. Injected volumes beyond 1 PV indicate that some of the scCO$_2$ was not retained within the sample, representing a leakage concern. 
We observe larger values of injected volume required to achieve similar levels of saturation with lower temperatures.
Fig. \ref{weightedSatvsTemp} explores the efficiency of the injection process by plotting the maximum weighted saturation of scCO$_2$ as a function of injected volume, representing the largest achievable saturation volume without scCO$_2$ breakthrough.
Data points closer to the diagonal represent maximum injection efficiency, and this trend improves with higher temperatures.   

\begin{suppfigure}[ht!]
  \includegraphics[width=\linewidth]{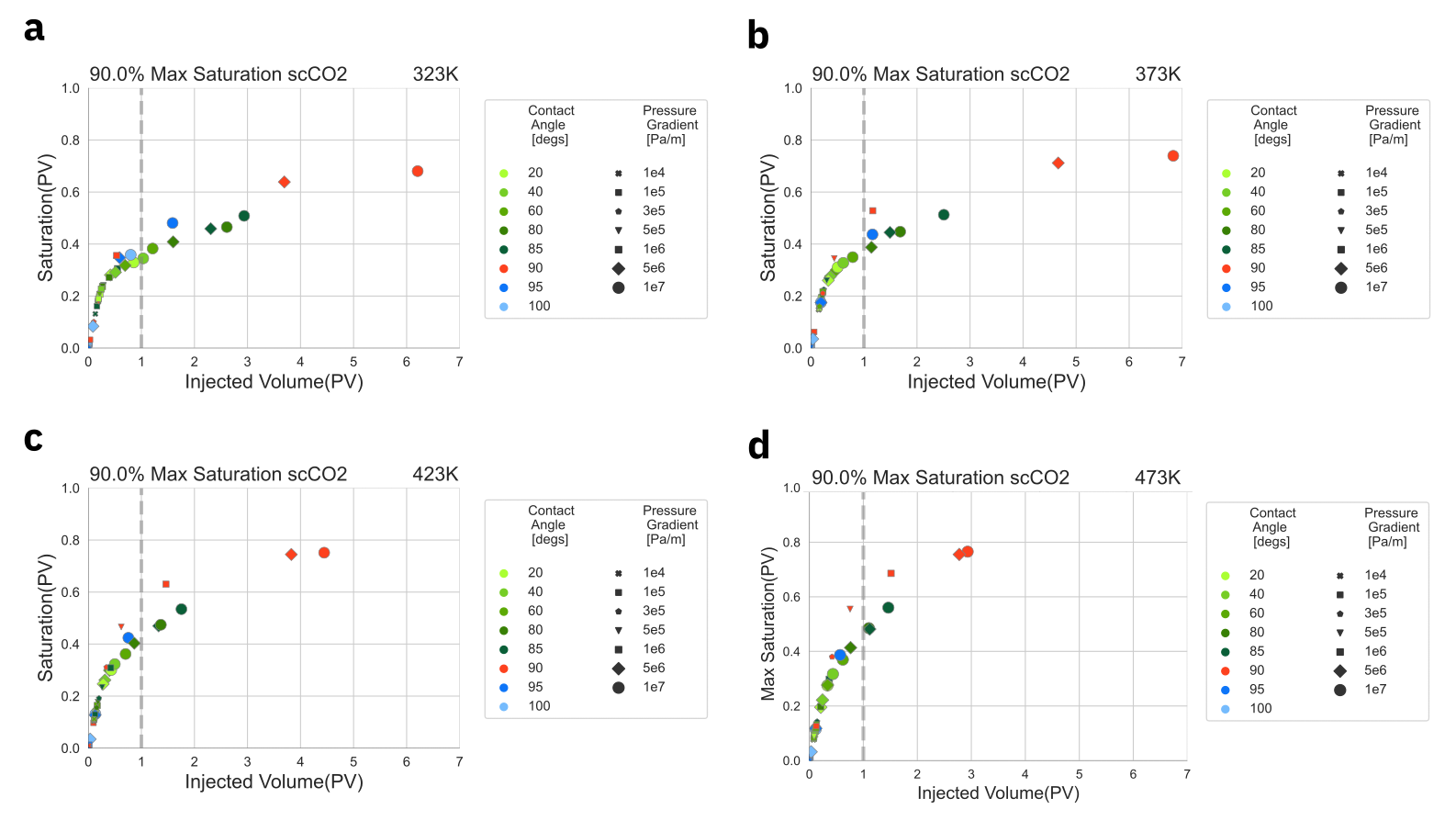}
  \caption{\textbf{Value at 90\% maximum saturation \textit{vs.} injected volume.} Green colors represents the CO$_2$-wet regime, red shows the zero capillary pressure regime and blue represents the intermediate-wet regime. Larger marker sizes represent higher pressure gradients at temperatures a) 323 K, b) 373 K c) 423 K and d) 473 K.}
  \label{MaxSatvsinjVolvsTemp}
\end{suppfigure}

\begin{suppfigure}[ht!]
  \includegraphics[width=\linewidth]{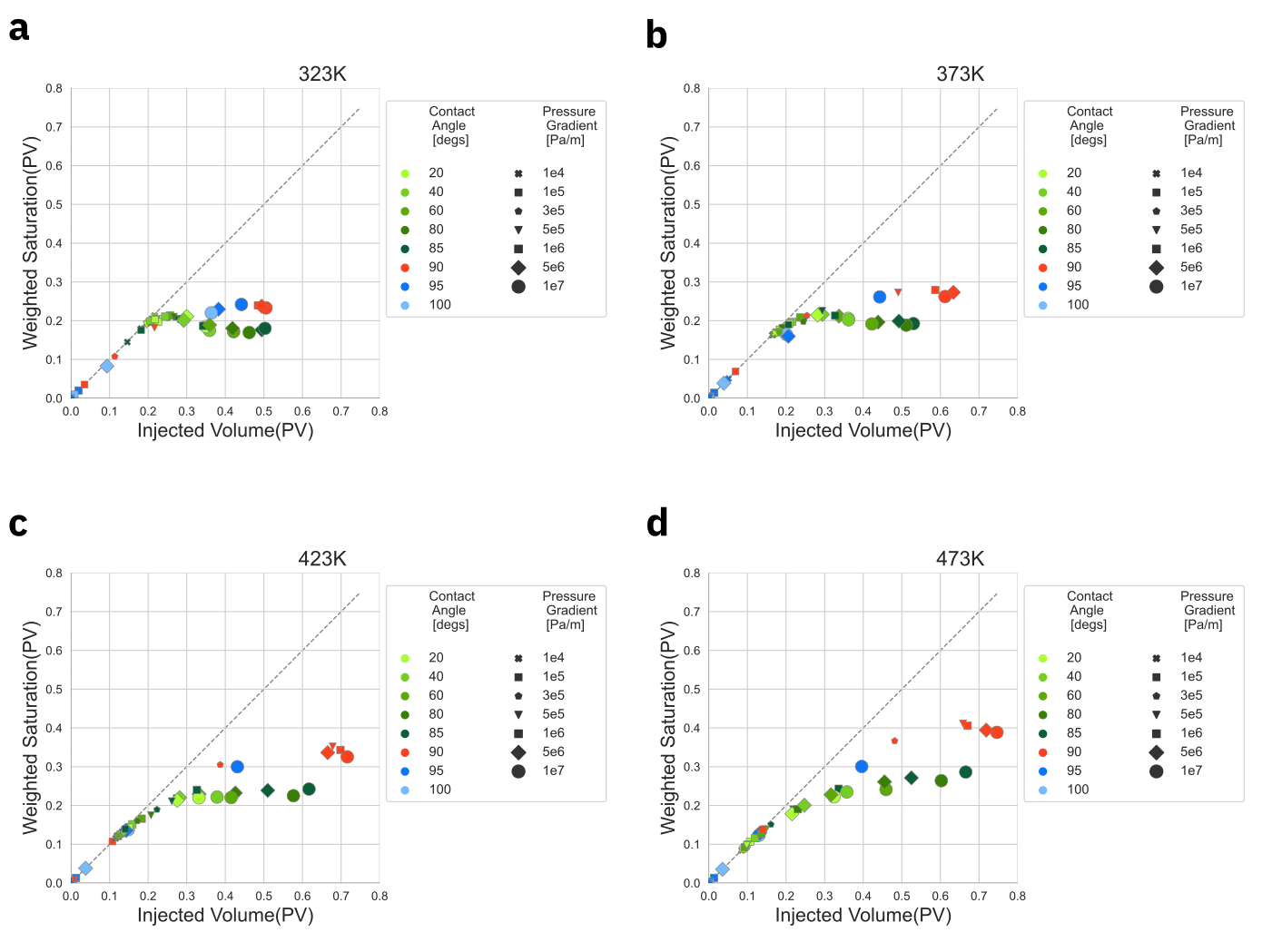}
  \caption{\textbf{Maximum weighted saturation vs injected volume.} Green colors represents a CO$_2$-wet regime, red shows the case of zero capillary pressure and, in blue, the intermediate-wet regime. Larger markers represent higher pressure gradients at temperatures a) 323 K, b) 373 K c) 423 K and d) 473 K.}
  \label{weightedSatvsTemp}
\end{suppfigure}

\end{additional}

\end{document}